\documentclass[prd,nofootinbib,twocolumn,showpacs,preprintnumbers,floatfix,amsmath,amssymb,amsfonts]{revtex4}

\usepackage{graphicx}
\usepackage{amsmath}
\usepackage{dcolumn}
\usepackage{epsfig}
\usepackage{psfrag}
\usepackage{lineno}

\RequirePackage{xspace}

\DeclareGraphicsExtensions{.epsi,.eps,.ps,.eps.gz,.ps.gz,.gif,.epsi.gz}
\newcommand{\e}      [1]   { {\ensuremath{ \times 10^{ {#1} } }}}

\def\vckm       {\ensuremath{ {V}_{CKM}}}

\def\theckmmatrix  {\ensuremath{ \left( \begin{array}{ccc} \vud & \vus & \vub \\ \vcd & \vcs & \vcb \\ \vtd & \vts & \vtb \end{array}\right).}}
\newcommand\vcq {\ensuremath{V_{cq}}}
\newcommand\vuq {\ensuremath{V_{uq}}}

\def\CPV {\ensuremath{CPV}}

\def\rhoz      {\ensuremath{\rho^{0}}}

\def\piz {\ensuremath{\pi^0}}

\RequirePackage{xspace}
\usepackage{color}
\definecolor{Red}{rgb}{1,0,0}
\definecolor{Magenta}{rgb}{0.5,0,0.5}
\definecolor{Blue}{rgb}{0,0,1}
\definecolor{Green}{rgb}{0,1,0}

\def\belle{Belle\xspace}
\def\belletwo{Belle II\xspace}

\usepackage{relsize}
\def\babar{\mbox{\slshape B\kern-0.1em{\smaller A}\kern-0.1em
    B\kern-0.1em{\smaller A\kern-0.2em R}}\xspace}

\def\superb{\ensuremath{\mathrm{Super}B}\xspace}

\def\lhcb {LHCb\xspace}

\def\epem       {\ensuremath{e^+e^-}\xspace}

\def\piz   {\ensuremath{\pi^0}\xspace}

\def\pip   {\ensuremath{\pi^+}\xspace}
\def\pim   {\ensuremath{\pi^-}\xspace}

\def\Kbar  {\kern 0.2em\overline{\kern -0.2em K}{}\xspace}

\def\Kz    {\ensuremath{K^0}\xspace}
\def\Kzb   {\ensuremath{\Kbar^0}\xspace}
\def\KzKzb {\ensuremath{\Kz \kern -0.16em \Kzb}\xspace}
\def\Kp    {\ensuremath{K^+}\xspace}
\def\Km    {\ensuremath{K^-}\xspace}

\def\KpKm  {\ensuremath{\Kp \kern -0.16em \Km}\xspace}
\def\KS    {\ensuremath{K^0_{\scriptscriptstyle S}}\xspace}
\def\KL    {\ensuremath{K^0_{\scriptscriptstyle L}}\xspace}

\def\Dbar    {\kern 0.2em\overline{\kern -0.2em D}{}\xspace}

\def\Dz      {\ensuremath{D^0}\xspace}
\def\Dzb     {\ensuremath{\Dbar^0}\xspace}
\def\DzDzb   {\ensuremath{\Dz {\kern -0.16em \Dzb}}\xspace}
\def\Dp      {\ensuremath{D^+}\xspace}
\def\Dm      {\ensuremath{D^-}\xspace}

\def\DpDm    {\ensuremath{\Dp {\kern -0.16em \Dm}}\xspace}

\def\B       {\ensuremath{B}\xspace}
\def\D       {\ensuremath{D}\xspace}
\def\Bbar    {\kern 0.18em\overline{\kern -0.18em B}{}\xspace}

\def\Bz      {\ensuremath{B^0}\xspace}
\def\Bzb     {\ensuremath{\Bbar^0}\xspace}
\def\BzBzb   {\ensuremath{\Bz {\kern -0.16em \Bzb}}\xspace}
\def\Bu      {\ensuremath{B^+}\xspace}
\def\Bub     {\ensuremath{B^-}\xspace}

\def\BpBm    {\ensuremath{\Bu {\kern -0.16em \Bub}}\xspace}
\def\Bs      {\ensuremath{B_s}\xspace}
\def\Bsb     {\ensuremath{\Bbar_s}\xspace}
\def\BsBsb   {\ensuremath{\Bs {\kern -0.16em \Bsb}}\xspace}

\def\BorBbar    {\kern 0.18em\optbar{\kern -0.18em B}{}\xspace}
\def\DorDbar    {\kern 0.18em\optbar{\kern -0.18em D}{}\xspace}
\def\KorKbar    {\kern 0.18em\optbar{\kern -0.18em K}{}\xspace}

\mathchardef\Upsilon="7107

\def\FourS {\ensuremath{\Upsilon{(4S)}}\xspace}
\def\FiveS {\ensuremath{\Upsilon{(5S)}}\xspace}

\mathchardef\Deltares="7101
\mathchardef\Xi="7104
\mathchardef\Lambda="7103
\mathchardef\Sigma="7106
\mathchardef\Omega="710A

\def\Deltabar{\kern 0.25em\overline{\kern -0.25em \Deltares}{}\xspace}
\def\Lbar{\kern 0.2em\overline{\kern -0.2em\Lambda\kern 0.05em}\kern-0.05em{}\xspace}
\def\Sigbar{\kern 0.2em\overline{\kern -0.2em \Sigma}{}\xspace}
\def\Xibar{\kern 0.2em\overline{\kern -0.2em \Xi}{}\xspace}
\def\Obar{\kern 0.2em\overline{\kern -0.2em \Omega}{}\xspace}
\def\Nbar{\kern 0.2em\overline{\kern -0.2em N}{}\xspace}
\def\Xb{\kern 0.2em\overline{\kern -0.2em X}{}\xspace}

\iffalse

\else

\fi

\newcommand{\tev}{\ensuremath{\mathrm{\,Te\kern -0.1em V}}\xspace}
\newcommand{\gev}{\ensuremath{\mathrm{\,Ge\kern -0.1em V}}\xspace}
\newcommand{\mev}{\ensuremath{\mathrm{\,Me\kern -0.1em V}}\xspace}
\newcommand{\kev}{\ensuremath{\mathrm{\,ke\kern -0.1em V}}\xspace}
\newcommand{\ev}{\ensuremath{\mathrm{\,e\kern -0.1em V}}\xspace}
\newcommand{\gevc}{\ensuremath{{\mathrm{\,Ge\kern -0.1em V\!/}c}}\xspace}
\newcommand{\mevc}{\ensuremath{{\mathrm{\,Me\kern -0.1em V\!/}c}}\xspace}
\newcommand{\gevcc}{\ensuremath{{\mathrm{\,Ge\kern -0.1em V\!/}c^2}}\xspace}
\newcommand{\mevcc}{\ensuremath{{\mathrm{\,Me\kern -0.1em V\!/}c^2}}\xspace}

\def\invpb {\ensuremath{\mbox{\,pb}^{-1}}\xspace}

\def\invfb   {\ensuremath{\mbox{\,fb}^{-1}}\xspace}
\def\invab   {\ensuremath{\mbox{\,ab}^{-1}}\xspace}

\def\mus  {\ensuremath{\rm \,\mus}\xspace}

\def\mus        {\ensuremath{\,\mu{\rm s}}\xspace}    

\def\to                 {\ensuremath{\rightarrow}\xspace}

\def\pep2{PEP-II}

\def\gsim{{~\raise.15em\hbox{$>$}\kern-.85em
          \lower.35em\hbox{$\sim$}~}\xspace}
\def\lsim{{~\raise.15em\hbox{$<$}\kern-.85em
          \lower.35em\hbox{$\sim$}~}\xspace}

\newcommand\vud {\ensuremath{V_{ud}}}
\newcommand\vus {\ensuremath{V_{us}}}
\newcommand\vub {\ensuremath{V_{ub}}}
\newcommand\vcd {\ensuremath{V_{cd}}}
\newcommand\vcs {\ensuremath{V_{cs}}}
\newcommand\vcb {\ensuremath{V_{cb}}}
\newcommand\vtd {\ensuremath{V_{td}}}
\newcommand\vts {\ensuremath{V_{ts}}}
\newcommand\vtb {\ensuremath{V_{tb}}}
\def\vckm       {\ensuremath{ {V}_{CKM}}}

\def\theckmmatrix  {\ensuremath{ \left( \begin{array}{ccc} \vud & \vus & \vub \\ \vcd & \vcs & \vcb \\ \vtd & \vts & \vtb \end{array}\right)}}

\def\CP                {\ensuremath{C\!P}\xspace}
\def\CPT               {\ensuremath{C\!PT}\xspace} 

\def\P       {\ensuremath{P}\xspace}

\def\Vud  {\ensuremath{|V_{ud}|}\xspace}
\def\Vcd  {\ensuremath{|V_{cd}|}\xspace}
\def\Vtd  {\ensuremath{|V_{td}|}\xspace}
\def\Vus  {\ensuremath{|V_{us}|}\xspace}
\def\Vcs  {\ensuremath{|V_{cs}|}\xspace}

\def\Vub  {\ensuremath{|V_{ub}|}\xspace}
\def\Vcb  {\ensuremath{|V_{cb}|}\xspace}

\def\deltat{\ensuremath{{\rm \Delta}t}\xspace}

\xspace

\def\jetset74   {\mbox{\tt Jetset \hspace{-0.5em}7.\hspace{-0.2em}4}\xspace}

\def\alphameasUTFit{\ensuremath{(91.4\pm 6.1)^\circ}\xspace}
\def\betameasUTFit{\ensuremath{(21.1\pm 0.9)^\circ}\xspace}
\def\gammameasUTFit{\ensuremath{(74\pm 11)^\circ}\xspace}

\def\sst{\scriptscriptstyle}
\def\lamf{\ensuremath{\lambda_{\sst f}}\xspace}
\def\xd{\ensuremath{x}\xspace}
\def\yd{\ensuremath{y}\xspace}

\long\def\inst#1{\par\nobreak\kern 4pt\nobreak
    {\it #1}\par\vskip 10pt plus 3pt minus 3pt}

\begin{document}


{\pagestyle{empty}

\par\vskip 3cm

\title{
\Large \boldmath
Time-dependent \CP asymmetries in $D$ and $B$ decays
}

\author{A. J. Bevan}
\author{G. Inguglia}
\affiliation{Queen Mary, University of London, Mile End Road, E1 4NS, United Kingdom}

\author{B. Meadows}
\affiliation{University of Cincinnati, Cincinnati, Ohio 45221, USA}

\date{\today}

\preprint{UCHEP-11-05}
\begin{abstract}
%
%
We examine measurements of time-dependent \CP asymmetries that could
be made in new and future flavour facilities.  In charm decays, where they
can provide a unique insight into the flavor changing structure of 
the Standard Model, we examine a number of decays to \CP eigenstates
and describe a framework that can be used to interpret the measurements.
Such measurements can provide a precise determination of the charm
mixing phase, as well as constraints on the Standard Model description of CP violation
an possible new physics contributions.
We make a preliminary assessment, based on statistical considerations, of
the relative capabilities of LHCb with data from $pp$ collisions, with
\belletwo and \superb using data from $B_d$, $B_s$ and charm thresholds.
We discuss the measurements required to perform direct and indirect 
tests of the charm unitarity triangle and its relationship with the 
usual $B_d$ triangle.  We find that, while theoretical and experimental 
systematic uncertainties may limit their interpretation, useful
information on the unknown charm mixing phase, and on the possible 
existence of new physics can be obtained.
We point out that, for $B_d$ decays, current experimental bounds on 
$\Delta\Gamma_{B_d}$ will translate into a significant systematic uncertainty 
on future measurements of $\sin 2\beta$ from $b \to c\overline{c}s$ decays.  
The possibilities for simplified $B_s$ decay asymmetry measurements at \superb 
and \belletwo are also reviewed. 
\end{abstract}

\pacs{13.25.Hw, 12.15.Hh, 11.30.Er}

\maketitle

\vfill

}

\setcounter{footnote}{0}

\section{Introduction}
\label{sec:intro}

The Standard Model (SM) description of quark mixing and \CP violation
is described by the Cabibbo-Kobayashi-Maskawa (CKM) matrix~\cite{Cabibbo:1963yz,Kobayashi:1973fv}.  This matrix
can be written as
\begin{eqnarray}
\vckm = \theckmmatrix,\label{eq:ckmmatrix}
\end{eqnarray}
where the $V_{ij}$ are coupling strengths for up-type to down 
type quark transitions.
Unitarity
of the CKM matrix gives rise to six triangles in a complex plane, one
of which, the $bd$ triangle, has been extensively studied by the \B Factories, and has 
earned the name of `The unitarity triangle'.  The unitarity triangle
is%
\footnote{We depart from the usual convention by defining the complex conjugates
of the triangle sides.}
\begin{eqnarray}
\vud^*\vub + \vcd^*\vcb + \vtd^*\vtb = 0. \label{eq:unitaritytriangle}  
\end{eqnarray}
The angles of the 
unitarity triangle are $\alpha=\alphameasUTFit$, $\beta = \betameasUTFit$,
and $\gamma=\gammameasUTFit$~\cite{utfit-website-2010,hfag}.  Given
that, for three generations, these angles must add up to $180^\circ$, the 
measurements of $\alpha$ and $\beta$ provide a strong constraint
on the value of $\gamma$ in the SM.  Precision tests of the CKM mechanism
have been made only for transitions of down type-quarks from the second and 
third generations, and time-dependent \CP asymmetries have only been 
measured in the third generation (\B decays).  In order to study the corresponding phenomena with an
up-type quark one has to study the charm decays as top quarks hadronize before
being able to form a quasi-stable meson. In this paper we concentrate
on examining approaches that may be usable, at existing and future experimental
facilities, in order to study time-dependent \CP asymmetries 
in the charm sector, and how results might be interpreted in the context of the 
CKM matrix.  In particular, we focus on ways in which the SM expectations of
the ``charm triangle'', defined in Eq.~(\ref{eq:charmtriangle}) might be examined
$-$ something that has yet to be done. 
In addition to this, we make a few observations on
measurements of $B_d$ and $B_s$ decays in Sections~\ref{sec:bdecays}
and \ref{sec:bsdecays}.

In addition to the $bd$ triangle, unitarity of the CKM matrix also gives rise 
to the charm ($cu$) triangle 
\begin{eqnarray}
\vud^*\vcd + \vus^*\vcs + \vub^*\vcb = 0, \label{eq:charmtriangle}  
\end{eqnarray}
which depends on the weak phase $\gamma$ by virtue of the presence of 
the factor \vub. The angles
of the charm triangle can be written as $\alpha_c$, $\beta_c$,
and $\gamma_c$.  Some time ago Bigi and Sanda~\cite{Bigi:1999hr} 
pointed out that $\gamma_c \simeq \gamma$, and 
$\alpha_c = 180^\circ-\gamma_c + {\cal O}(\lambda^4) = 180^\circ-\gamma + {\cal O}(\lambda^4)$
\footnote{Our nomenclature differs from that used in Ref.~\cite{Bigi:1999hr}.  Our
angles can be related to theirs as follows: $\beta_c \equiv \phi_3^{cu}$, 
$\gamma_c \equiv \phi_2^{cu}$, and $\alpha_c\equiv \phi_1^{cu}$.}.
Hence the charm and unitarity triangles are related in a simple way.
Using the Buras et al. variant of the Wolfenstein
parameterization~\cite{Wolfenstein:1983yz,Buras:1994ec} of the CKM matrix up to ${\cal O}(\lambda^5)$, the 
weak phase $\beta_c$ can be estimated to be $\sim 0.035^\circ$. 
Hence the sum of $\alpha_c$ and $\gamma_c$ should be essentially $180^\circ$.
Existing constraints on the Wolfenstein parameters can be used to 
give a clean prediction of the $cu$ triangle parameters.
In order to verify if the CKM matrix is the correct description of quark mixing the angles
$\alpha_c$, $\beta_c$, and $\gamma_c$ need to be measured, as well as the 
sides of this triangle.
The $e^+e^-$ collider experiment
\superb is the only facility where one can, in principle, perform all of the 
necessary measurements to perform a complete cross-check of the two triangles.
This requires large samples of $\B$, $\D$ and $\D_s$ mesons, which \superb
will accumulate through runs at charm threshold, and at the \FourS.
In order to interpret time-dependent measurements in terms of angles of the 
$cu$ triangle, a precise measurement of the charm mixing phase is required.
We propose that one studies $D\to K^+K^-$ to measure this mixing phase,
and the difference between measurements of $D\to K^+K^-$ and $D\to \pi^+\pi^-$
will then give $-2\beta_{c,eff}$.
A single measurement of, or constraint on, $\beta_c$, predicted to be $(0.0350 \pm 0.0001)^\circ$,
would clearly be of interest, but this will require a careful study of effects of other amplitudes and possible long range effects.  Any observed deviation from this expectation would then
be an indication of new physics (NP).
Indeed it is worth noting that the measurement of $\sin2\beta$ from $B$ meson decays 
to final states including Charmonium and a neutral kaon are inconsistent with 
the SM at the level of $3.2\sigma$~\cite{Lunghi:2011xy}.  This result
is a strong motivation to perform the corresponding studies of the SM in the charm 
sector, which is the subject of this paper.

The possibility of large \CP violation effects in charm decays has been
discussed elsewhere~\cite{Buccella:1994nf,Bianco:2003vb,Petrov:2004gs,Grossman:2006jg}, 
however until now these have focused on time-integrated
measurements, and ignored possible time-dependent effects.  It is clear that we need
precision experimental tests of the unitarity triangle, in particular the 
angle $\gamma$, and the sides of both the charm and unitarity triangles.  In addition to this 
we also need to start measuring the charm triangle angles precisely in order to validate
the CKM description of \CP violation for up-type quarks.  Theoretical uncertainties
will ultimately limit the constraints that can be placed on the SM, and we discuss
some of the issues here.  The remainder of this paper
outlines the details required to perform time-dependent \CP measurements
in the charm sector, using \CP eigenstate decays, and constraints on the sides of the charm triangle,
before making a few observations on $B_d$ and $B_s$ decays.

\begin{widetext}
\section{The CKM Matrix}
\label{sec:CKM}

The CKM matrix given in Eq.~(\ref{eq:ckmmatrix}) can be parameterized in a number
of different ways.  The Wolfenstein parameterization~\cite{Wolfenstein:1983yz} is 
an expansion in terms of $\lambda = \sin \theta_c$, $A$, $\rho$, and $\eta$, where $\theta_c$ 
is the Cabibbo angle.  A variant on this parameterization has been proposed Buras et al.~\cite{Buras:1994ec},
and has the advantage of preserving unitarity to all orders in $\lambda$ in the ``$bd$"
triangle, though possibly not in the ``$cu$" triangle.
The Buras et al. variant of the CKM matrix up
to and including terms ${\cal O}(\lambda^5)$ is given by
\begin{eqnarray}
\vckm = \left( 
 \begin{array}{ccc}
  1 - \lambda^2 / 2 -\lambda^4/8              &   \lambda                              & A\lambda^3 (\rho - i\eta) \\
  -\lambda +A^2\lambda^5[1 - 2(\rho+i\eta)]/2 & 1 - \lambda^2/ 2 - \lambda^4(1+4A^2)/8 & A \lambda^2  \\
  A\lambda^3[1-(1-\lambda^2/2)(\rho +i\eta)]  & -A \lambda^2 + A\lambda^4[1-2(\rho+i\eta)]/2   & 1 -A^2\lambda^4/2
 \end{array}
\right) + {\cal O}(\lambda^6).
\label{eq:ckmmatrixburas}
\end{eqnarray}
The choice of convention used to interpret data in terms of physical observables is 
irrelevant as long as sufficient
terms in the expansion are used.  Expansions to ${\cal O}(\lambda^3)$, have been 
sufficient for the \B factories era, however one should consider additional terms 
as we move into the era of LHCb and the Super Flavor Factories (SFF's).

The apex of the unitarity triangle, obtained when Eq.~(\ref{eq:unitaritytriangle}) is normalized by $\vcd\vcb^*$, 
is given by the coordinates $(\overline{\rho},\, \overline{\eta})$, where
\begin{eqnarray}
\overline{\rho} = \rho \left[ 1 - \lambda^2/2 + {\cal O}(\lambda^4)\right], \\
\overline{\eta} = \eta \left[ 1 - \lambda^2/2 + {\cal O}(\lambda^4)\right].
\end{eqnarray}
The CKM matrix may be written in terms of $\overline{\rho}$ and $\overline{\eta}$ as
\begin{eqnarray}
\vckm = \left(
 \begin{array}{ccc}
  1 - \lambda^2 / 2 -\lambda^4/8              &   \lambda                              & A\lambda^3 (\bar\rho - i\bar\eta)(1+\lambda^2/2) \\
  -\lambda +A^2\lambda^5[1 - 2(\bar\rho+i\bar\eta)]/2 & 1 - \lambda^2/ 2 - \lambda^4(1+4A^2)/8 & A \lambda^2  \\
  A\lambda^3[1-\bar\rho -i\bar\eta]  & -A \lambda^2 + A\lambda^4[1-2(\bar\rho+i\bar\eta)]/2   & 1 -A^2\lambda^4/2
 \end{array}
\right) + {\cal O}(\lambda^6).
\label{eq:ckmmatrixburasrhobaretabar}
\end{eqnarray}
\end{widetext}
Current constraints on the CKM parameters $A$, $\lambda$, $\rho$, $\eta$, $\overline{\rho}$, and $\overline{\eta}$
from global fits~\cite{utfit-website-2010,Hocker:2001xe} are given in Table~\ref{tbl:wolf}.

\begin{table}[!ht]
\caption{Constraints on the CKM parameters $A$, $\lambda$, $\rho$, $\eta$, $\overline{\rho}$, and $\overline{\eta}$ 
obtained by the UTFit and CKM fitter groups.}\label{tbl:wolf}
\begin{center}
\begin{tabular}{lccc}
Parameter         & UTFit & CKM Fitter & Mean Used \\ \hline \hline
$\lambda$         & $0.22545 \pm 0.00065$ & $0.22543 \pm 0.00077$
                  & $0.22544 \pm 0.00705$
\\
$A$               & $0.8095 \pm 0.0095$   & $0.812^{+0.013}_{-0.027}$
                  & $0.811  \pm 0.015$
\\
$\rho$            & $0.135 \pm 0.021$     & $-$
                  & $-$
\\
$\eta$            & $0.367 \pm 0.013$     & $-$ 
                  & $-$
\\
$\overline{\rho}$ & $0.132 \pm 0.020$     & $0.144\pm 0.025$
                  & $0.138 \pm 0.022$
\\
$\overline{\eta}$ & $0.358 \pm 0.012$     & $0.342^{+0.016}_{-0.015}$ 
                  & $0.350 \pm 0.014$
\\ \hline
\end{tabular}
\end{center}
\end{table}

The angles of the unitarity triangle given in Eq.~(\ref{eq:unitaritytriangle}) 
are $\alpha$, $\beta$, and $\gamma$, where%
\begin{eqnarray}
\alpha &=& \arg\left[-\vtd \vtb^*/\vud\vub^*\right] =\alphameasUTFit, \label{eq:alpha}\\
\beta  &=& \arg\left[ -\vcd\vcb^* / \vtd\vtb^*\right] =\betameasUTFit, \label{eq:beta}\\
\gamma &=& \arg\left[ -\vud\vub^* / \vcd\vcb^* \right] =\gammameasUTFit. \label{eq:gamma}
\end{eqnarray}
The most precisely measured angles are $\alpha$ and $\beta$ using $B$ meson
decays into $\rho\rho$~\cite{Aubert:rhorho,Somov:rhorho} and charmonium final states~\cite{Chen:2006nk,Aubert:2009yr}, respectively.
Given unitarity, in the SM with just three generations, only two of these angles are independent, hence $\gamma$ is, 
in principle, a redundant cross-check of the CKM matrix.  

%
These angles can also be computed from values and uncertainties for 
$A$, $\lambda$, $\bar\rho$ and $\bar\eta$.  Taking simple averages of CKM 
Fitter and UTFit values in Table~\ref{tbl:wolf}, the angles, computed to order 
$\lambda^6$, are 
\begin{eqnarray}
  \alpha &=& (89.4\pm 4.3)^{\circ},  \\
  \beta   &=& (22.1\pm 0.6)^{\circ}, \\
  \gamma&=& (68.4\pm 3.7)^{\circ}.
\end{eqnarray}

Comparing Eqns~(\ref{eq:beta}) and (\ref{eq:gamma}) with 
Eq.~(\ref{eq:ckmmatrixburas}), 
one can see that $\vtd \simeq \Vtd e^{-i\beta}$, and $\vub \simeq \Vub e^{-i\gamma}$.
These relations are exact for low orders of $\lambda$, and
the equality breaks down as \vcd\ is complex at order $\lambda^5$.

The angles of the charm unitarity triangle given in Eq.~(\ref{eq:charmtriangle}) are
\begin{eqnarray}
\alpha_c &=& \arg\left[- \vub^*\vcb/ \vus^*\vcs \right], \label{eq:alphac}\\
\beta_c  &=& \arg\left[- \vud^*\vcd / \vus^*\vcs \right], \label{eq:betac}\\
\gamma_c &=& \arg\left[- \vub^*\vcb / \vud^*\vcd\right], \label{eq:gammac}
\end{eqnarray}
where as already noted $\gamma_c \simeq \gamma$ and 
$\alpha_c = 180^\circ - \gamma + {\cal O}(\lambda^4)$.  
%
Again, using the averages of CKM Fitter and UTFit values for $A$, $\lambda$, $\bar\rho$ and $\bar\eta$ and their errors, we predict that, to order $\lambda^6$
\begin{eqnarray}
  \alpha_c &=& (111.5\pm 4.2)^{\circ}, \\
  \beta_c &=& (0.0350\pm 0.0001)^{\circ}, \\
  \gamma_c &=& (68.4\pm 0.1)^{\circ}.
\end{eqnarray}

\begin{figure}[!ht]
\begin{center}
  \resizebox{8cm}{!}{
\includegraphics{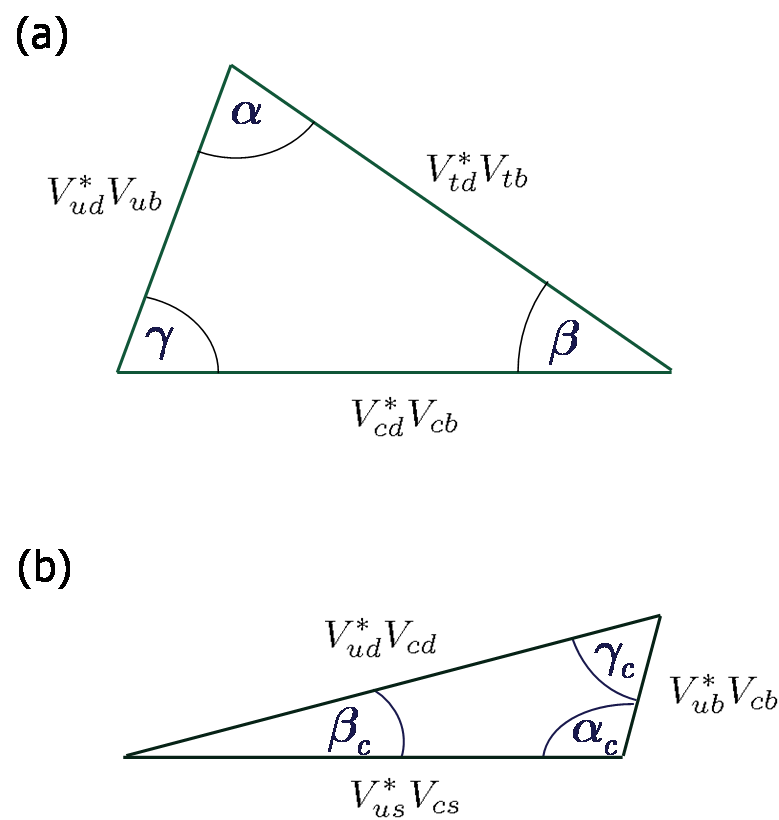}
}
  \caption{(top) The $bd$ unitarity triangle in Eq.~(\ref{eq:unitaritytriangle}), and (bottom) the $cu$
unitarity triangle of Eq.~(\ref{eq:charmtriangle}).}
  \label{fig:triangles}
\end{center}
\end{figure}

These predictions for the angles of the charm triangle could, and should, be 
tested experimentally, either directly (through time-dependent \CP asymmetries) 
or indirectly (through measurements of the sides of the triangle).
On comparing Eq.~(\ref{eq:betac}) with Eq.~(\ref{eq:ckmmatrixburas}), one can 
see that $\vcd = \Vcd e^{i(\beta_c-\pi)}$.
Both the $bd$ and $cu$ triangles are shown in Fig.~\ref{fig:triangles}.

\section{Time-dependent evolution}
\label{sec:tdep}

Neutral meson mixing is a phenomenon that only occurs for $K$, $D$, and $B_{d,s}$ 
mesons (Charge conjugation is implied throughout). 
Here, in describing the formalism common to these systems, we refer to the 
mesons as $P$.  The effective Hamiltonian describing neutral meson mixing 
is given by 
\begin{eqnarray}
{\cal H}_{eff} &=& {\bf M} - \frac{i {\bf \Gamma}}{2},\label{eq:tdep:heff}\\
               &=& \left( \begin{array}{cc} M_{11} & M_{12} \\ M_{21} & M_{22} \end{array}\right)
                     - \frac{i}{2}\left( \begin{array}{cc} \Gamma_{11} & \Gamma_{12} \\ \Gamma_{21} & \Gamma_{22} \end{array}\right).
\end{eqnarray}
Hence neutral meson mixing can be described by
\begin{eqnarray}
i\frac{\partial }{\partial t}
\left( \begin{array}{c} |P^0\rangle \\ |\overline{P}^0\rangle \end{array} \right)
= \left(M-\frac{i}{2}\Gamma\right)
\left( \begin{array}{c}|P^0\rangle \\ |\overline{P}^0\rangle \end{array} \right),\label{eq:tdep:heffmatrix}
\end{eqnarray}
where $|P^0\rangle$ and $|\overline{P}^0\rangle$ are strong eigenstates of 
neutral $B$, $D$, or $K$ mesons.  
The matrix elements in Eq.~(\ref{eq:tdep:heffmatrix}) must satisfy 
$M_{11} = M_{22}$ and $\Gamma_{11} = \Gamma_{22}$ in order to be consistent with 
\CPT symmetry.   A further constraint can be obtained in the limit of \CP or $T$
invariance, where $\Gamma_{12}/M_{12}=\Gamma_{21}/M_{21}$ must be a real quantity.

One can write the mass eigenstates as an admixture of the strong eigenstates in the following 
way
\begin{eqnarray}
|P_{1,2}\rangle = p|P^0\rangle \pm q|\overline{P}^0\rangle,
\label{eq:tdep:admixture}
\end{eqnarray}
where $q^2+p^2=1$ to normalize the wave function, and 
\begin{eqnarray}
\frac{q}{p} = \sqrt{ \frac{M_{12}^* - i\Gamma_{12}^* / 2} {M_{12} - i\Gamma_{12} / 2} }.
\end{eqnarray}
The magnitude of $q/p$ is very nearly one in the SM.
If one considers the mass eigenstates under the \CP operator, it follows that $|P_1\rangle$
is \CP even, and $|P_2\rangle$
is \CP odd.  The mass and width differences $\Delta M$ and $\Delta \Gamma$ between the 
mass eigenstates are given by
\begin{eqnarray}
\Delta M &=& M_2 - M_1,\\
\Delta \Gamma &=& \Gamma_1 - \Gamma_2,
\end{eqnarray}
where neutral mesons oscillate from particle to anti-particle state with the characteristic
mixing frequency $\Delta M$.  Detailed discussions of this formalism can be found in
a number of text books.

\begin{widetext}
\subsection{Uncorrelated meson production}
\label{sec:tdepuncorrelated}

It can be shown that the general form of the time-evolution of a neutral meson decaying into
some final state $f$ is given by
\begin{eqnarray}
\Gamma(P^0 \to f) \propto e^{-\Gamma_1 t}\left[ \frac{\left(1+e^{\Delta\Gamma t} \right)}{2} + \frac{Re(\lamf)}{1+|\lamf|^2}\left(1-e^{\Delta\Gamma t} \right) + e^{\Delta\Gamma t / 2}\left(\frac{1-|\lamf|^2}{1+|\lamf|^2}\cos \Delta M t - \frac{2 Im(\lamf)}{1+|\lamf|^2}\sin \Delta M t  \right) \right],  \label{EQ:p0toffinal}\\
\Gamma(\overline{P}^0 \to f) \propto e^{-\Gamma_1 t}\left[ \frac{\left(1+e^{\Delta\Gamma t} \right)}{2} + \frac{ Re(\lamf)}{1+|\lamf|^2}\left(1-e^{\Delta\Gamma t} \right) + e^{\Delta\Gamma t / 2}\left(-\frac{1-|\lamf|^2}{1+|\lamf|^2}\cos \Delta M t + \frac{2 Im(\lamf)}{1+|\lamf|^2}\sin \Delta M t  \right) \right],\label{EQ:p0bartoffinal}
\end{eqnarray}
\end{widetext}
where 
\begin{eqnarray}
\lamf = \frac{q}{p} \frac{\overline{A}}{A},\label{eq:lambda}
\end{eqnarray}
and $A$ ($\overline{A}$) is the amplitude for the $P$ ($\overline{P}$) decay to a final state
$f$.  Note that $\lamf$ is not related to the CKM expansion parameter 
$\lambda$ discussed above, but is a complex parameter related to mixing and decay 
transitions.  The
time $t=0$ is defined by the production of a definite meson state (flavor, CP or mixed flavor), that subsequently 
evolves as a $P-\overline{P}$ admixture until it too decays.  The identification of the flavor of 
a meson state at some fixed point in time is critical for a time-dependent measurement and is 
discussed in Section~\ref{sec:ftag}. If $|q/p| \neq 1$ then there is \CP violation in mixing, and 
if $|A|^2 \neq |\overline{A}|^2$ there is direct \CP violation, hence a measurement of 
the real and imaginary parts of $\lamf$ (or equivalently the magnitude and phase) is able to 
probe the combination of these two effects i.e. interference between mixing and decay.
It should be noted that for all time-dependent \CP asymmetry measurements of $B_d^0$ decays made by experiments
until now the assumption that $\Delta \Gamma = 0$ has been used.  We discuss the ramifications 
of this assumption in Section~\ref{sec:bdecays}.

A time-dependent decay rate asymmetry can be computed from Eqns~(\ref{EQ:p0toffinal}) and (\ref{EQ:p0bartoffinal}) 
as follows
\begin{eqnarray}
 {\cal A}(t) &=& \frac{ \overline{\Gamma} (t) - \Gamma (t) } {\overline{\Gamma}(t) + \Gamma(t) },\label{eq:asym}\\
 &=& 2 e^{\Delta \Gamma t/2} \frac{ (|\lamf|^2 - 1)\cos \Delta M t + 2Im \lamf \sin\Delta M t}{(1 + |\lamf|^2)(1+e^{\Delta \Gamma t}) + 2 Re\lamf ( 1 - e^{\Delta\Gamma t})}, \nonumber
\end{eqnarray}
where $\overline{\Gamma} (t)$ and $\Gamma (t)$ are the time-dependent rates, respectively, 
for $\overline{P}^0 \to \overline{f}$ and $P^0 \to f$ transitions.
The asymmetry depends on the real and imaginary parts of $\lamf$ as well as $|\lamf|^2$, hence 
it is possible to extract $\lamf$ from data in terms of only two parameters, the real and imaginary parts of $\lamf$,
as $|\lamf|^2$ is completely correlated with those parameters.  
This formalism is normally written in terms of hyperbolic functions, and one can derive those results
by combining the exponential factors in the equations above.
In the limit that 
$\Delta \Gamma = 0$, Eq.~(\ref{eq:asym}) reduces to the familiar result
\begin{eqnarray}
 {\cal A}(t) &=& -C \cos \Delta M t + S \sin\Delta M t,
\end{eqnarray}
where
\begin{eqnarray}
 S &=& \frac{2Im \lamf}{1 + |\lamf|^2}, \text{ and } C = \frac{1 - |\lamf|^2 }{1 + |\lamf|^2}.\label{eq:bzcpparameters}
\end{eqnarray}
This approximation has been used in the $B$ factory measurements of the 
angles in the $bd$ unitarity triangle.  For future measurements, we note,
the validity of the assumption that $\Delta\Gamma=0$ will need further
checking.

\subsection{Correlated production of neutral mesons}
\label{sec:tdepcorrelated}

Neutral $K$, $D$, or $B$ mesons are produced in correlated pairs in $e^+e^-$ collections with center of mass 
energies corresponding to the $\phi$, $\psi(3770)$, or \FourS resonances, respectively.
The time-dependence of
such mesons is complicated by the issue that the pairs of neutral mesons are produced in a coherent wave function
consisting of exactly one $|P^0\rangle$ and one $|\overline{P}^0\rangle$ state until one of the mesons decays
and the correlated wave function collapses.  At that point in time $t_1$, the second $P$ meson starts to oscillate
with mixing frequency $\Delta M$, until eventually this also decays at some later time $t_2$.  The time-difference
\deltat between these two meson decays replaces the variable $t$ used to describe the evolution of uncorrelated mesons.
The sign of \deltat is taken to be the difference between the decay time of a meson into a \CP eigenstate minus
that of the decay into a flavor specific final state (See Section~\ref{sec:ftag}).  Hence events where
the \CP eigenstate decay is the second one to occur have positive values of \deltat, and those where
the \CP eigenstate decay occurs first have negative values of \deltat.

\begin{widetext}
The corresponding time-dependence is given by
\begin{eqnarray}
\Gamma(P^0 \to f) &\propto& e^{-\Gamma_1 |\deltat|}\left[ \frac{h_+}{2} + \frac{ Re(\lamf)}{1+|\lamf|^2}h_- + e^{\Delta\Gamma \deltat / 2}\left(\frac{1-|\lamf|^2}{1+|\lamf|^2}\cos \Delta M \deltat - \frac{2 Im(\lamf)}{1+|\lamf|^2}\sin \Delta M \deltat  \right) \right], \label{EQ:p0toffinaldeltat}\\
\Gamma(\overline{P}^0 \to f) &\propto& e^{-\Gamma_1 |\deltat|}\left[ \frac{h_+}{2} + \frac{ Re(\lamf)}{1+|\lamf|^2}h_- + e^{\Delta\Gamma \deltat / 2}\left(-\frac{1-|\lamf|^2}{1+|\lamf|^2}\cos \Delta M \deltat + \frac{2 Im(\lamf)}{1+|\lamf|^2}\sin \Delta M \deltat  \right) \right],\label{EQ:p0bartoffinaldeltat}
\end{eqnarray}
where
\begin{eqnarray}
h_{\pm} =  1 \pm e^{\Delta\Gamma \deltat}.
\end{eqnarray}
Hence the time-dependent \CP asymmetry becomes
\begin{eqnarray}
 {\cal A}(\deltat) &=& \frac{ \overline{\Gamma} (\deltat) - \Gamma (\deltat) } {\overline{\Gamma}(\deltat) + \Gamma(\deltat) } =
 2 e^{\Delta \Gamma \deltat/2} \frac{ (|\lamf|^2 - 1)\cos \Delta M \deltat + 2Im \lamf \sin\Delta M \deltat}{(1 + |\lamf|^2) h_+ + 2 h_- Re\lamf} 
 \label{eq:asymdeltat}
\end{eqnarray}
\end{widetext}
and is similar to that for uncorrelated $\P^0$ production.  In this case, 
however, at $\deltat=0$, the two $P$'s are completely correlated\footnote{E.g., if the first decays to a $\CP=-1$ eigenstate then, at 
$\deltat=0$, the other has $\CP=+1$ and no odd-\CP components will appear in 
its own decay.}
so that the decay of either one is ``filtered" by the decay mode of the 
other.  When $\Delta\Gamma = 0$, $h_+ = 2$ and $h_-=0$.

\begin{figure}[!ht]
\begin{center}
  \resizebox{8cm}{!}{\includegraphics{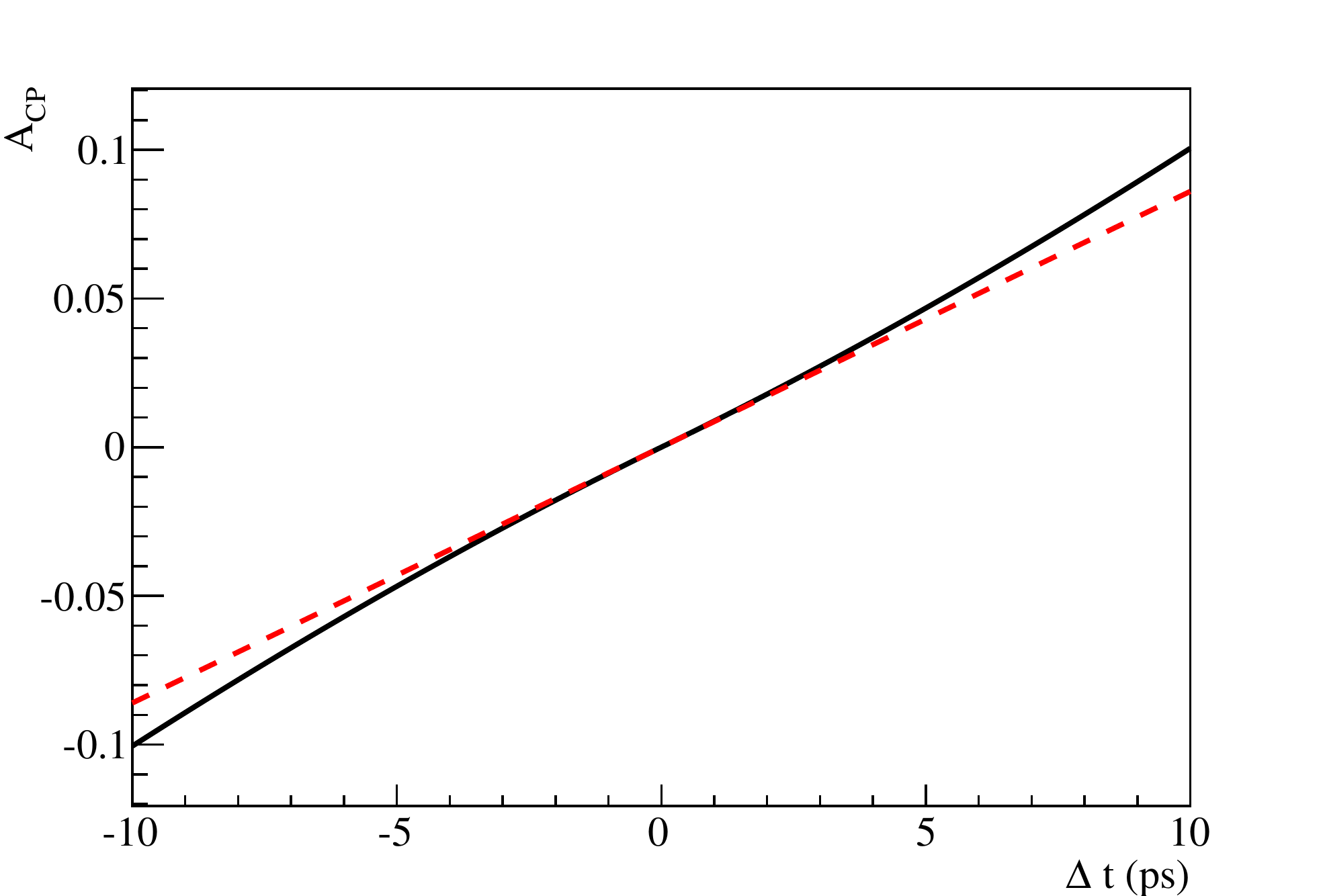}}
  \resizebox{8cm}{!}{\includegraphics{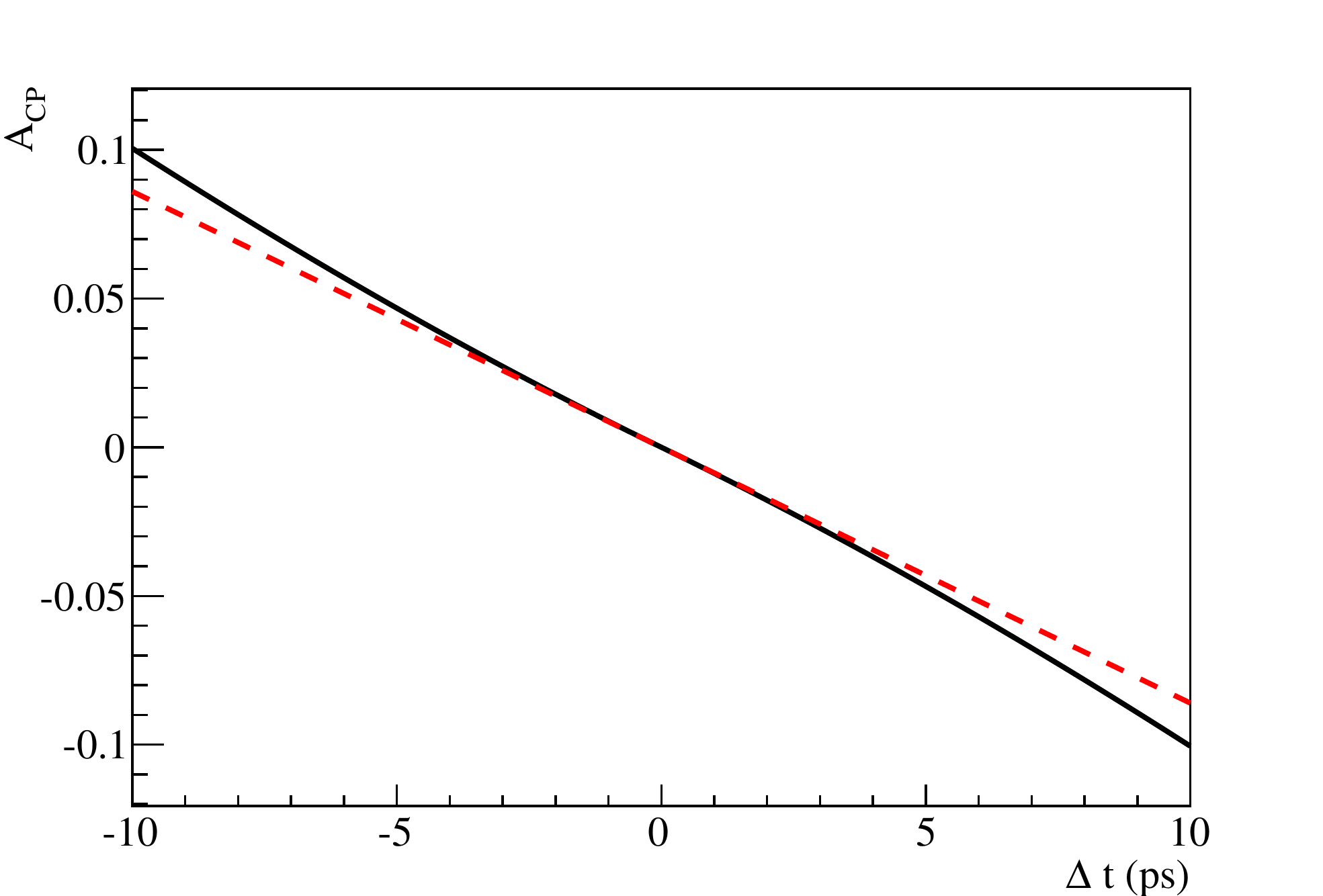}}
  \caption{Distribution of ${\cal A}(\deltat)$
   for \Dz mesons with (top) $Re\lamf = Im\lamf = 1/\sqrt{2}$ and $\CP=+1$
   and (bottom) the expected asymmetry for $\CP=-1$ decay with the same value of \lamf.
   These distributions assume $q/p=1$, and the solid (dashed) line corresponds to 
   $y=0.01$ ($0.00$).
   For $B_d$ decays, 1.5 full sinusoidal oscillations are observed in the time 
   interval presented here.}
  \label{fig:dtimedep}
\end{center}
\end{figure}

For charm decays the measured parameters normally used are $x$ and $y$
(or a pair of variables related to $x$ and $y$ by a simple rotation), 
where
\begin{eqnarray}
x = \frac{\Delta M}{\Gamma}, \text{ and } y = \frac{\Delta \Gamma}{2\Gamma}.
\end{eqnarray}
Current experimental constraints \cite{hfag} give $x\sim 0.005$ and $y\sim 0.01$.
In order to illustrate Eq.~(\ref{eq:asymdeltat}), the 
distribution ${\cal A}(\deltat)$ for $\Dz$ decays assuming 
$Re\lamf = Im\lamf = 1/\sqrt{2}$ is shown in Fig.~\ref{fig:dtimedep}
using $x= 0.005$ and $y= 0.01$~\cite{hfag}.  It is clear from this illustration
that oscillations in the charm sector are slow compared with those 
from $B_d$ or $B_s$ decays, and the \CP asymmetry varies almost linearly with
$\deltat$.  While an asymmetry is observable, one
will require large statistics to be accumulated in order to make 
a non-trivial measurement.  It should also be noted from 
Eq.~(\ref{eq:asymdeltat}) that precise knowledge of both $\Delta \Gamma$
and $\Delta M$ will be required in order to translate the slope of the
asymmetry into a constraint on $\lamf$ for a given decay channel, as indicated
by the two curves shown in Fig.~\ref{fig:dtimedep}.

\subsection{Plausibility of measuring time-dependent \CP asymmetries.}

We will be considering three current or planned experimental scenarios where 
measurements of time-dependence of \CP asymmetry in charm decays 
might be possible.  These correspond to a) LHCb, b) a 2nd generation  
SFF - either \superb~\cite{O'Leary:2010af,Grauges:2010fi,Biagini:2010cc} or \belletwo~\cite{Abe:2010sj,Aushev:2010bq} running $\epem$ collisions at the $\Upsilon(4S)$ or c) Charm threshold - \superb running at the $\psi(3770)$ ($\Dz\Dzb$ threshold where the $D$'s are produced in a coherent state).

Event yields in each scenario will depend on the specific final states 
considered.  Estimates can be made based on the proven performance of 
\babar and \belle and of the current performance of LHCb
\cite{LHCb},
assuming that the current trigger efficiencies can be maintained and 
that the cross section will increase by a factor two for an energy
increase to 14~TeV from the current 7~TeV.  For the decays to \CP
eigenstates ($\Dz\to\Kp\Km$ or $\Dz\to\pip\pim$, for example) the yield 
from a 5 \invfb sample at LHCb is likely to be comparable to that 
expected from each of the SFF's.
An upgrade to LHCb could provide a factor 10 more events.  Background 
levels at LHCb are, however, considerably larger than those anticipated
at a SFF.  For modes with higher multiplicity, lower trigger
efficiencies will probably contribute to LHCb sample sizes that are less 
competitive with the SFF's, again with larger background levels.  Event 
yields at charm threshold will be lower, since the \superb luminosity
is expected to be smaller at this energy by a factor 10.  Backgrounds, 
however, will be lower than at the $\Upsilon(4S)$.

To be able to measure $t$ or $\deltat$, $D$ mesons must be
produced in flight in the laboratory frame of reference.  This means 
that the flight length of the neutral $D$ mesons under study needs to 
exceed the detector resolution associated with reconstructing each 
final state studied.  We see, however, from Fig.~\ref{fig:dtimedep}, 
that the asymmetry varies almost linearly with decay time so, in 
principle, asymmetries need only be measured in a few regions of 
$\deltat$, perhaps even for just the two regions $\deltat<0$ and 
for $\deltat>0$.

LHCb should have decay time resolution that is superior to either of 
the SFF environments.  The \Dz's at LHCb are produced with 
momenta of hundreds of GeV/c and the time resolution is generally quite
small compared to the \Dz lifetime.  The $\Dz$ mesons at LHCb are 
produced both promptly and in $B$ decays, so care is needed to treat 
each separately.  The LHCb trigger has an efficiency that varies with 
decay length, and a data driven way to measure this variation is 
required if a systematic limit in precision is to be avoided.  
Certainly, a finer granularity in the time-dependence of any 
observed \CP asymmetry is surely possible in the LHCb than in 
either SFF environment.

Time resolution is more of an issue in the SFF environments.  
However, prompt $\Dz$'s from $\epem$ continuum can be cleanly 
distinguished from those from $B$ decay by applying a kinematic 
cut in momentum.  Also, event selection is not based on decay time, 
so that efficiency does not depend upon decay time.  

\Dz's produced in $\Upsilon(4S)$ decays have decay lengths dominated 
by the break-up momentum they acquire.  Assuming the performances 
of \superb and \belletwo are comparable, respectively, to those of 
\babar and \belle, the decay time resolution for these \Dz's are 
expected to be of order one half the \Dz lifetime.  This was 
sufficient for both \belle and \babar to observe mixing in the 
\Dz-\Dzb system.  Therefore, with the yields at the \superb 
expected to be at least two orders of magnitude greater, with
backgrounds that are similarly small, and with proven data driven 
techniques to estimate charge asymmetry effects, observation 
of \CP asymmetries and their time-dependence are at least 
conceivable.

At charm threshold, where coherent $\Dz$ pairs come from decays of
the $\psi(3770)$, the break-up momentum is
small so that measurements of $\Delta t$ rely upon the
boost from the asymmetric operation of \superb.  At the
$\Upsilon(4S)$, the boost ($\beta\gamma\sim 0.23$) is such that 
the decay length (one lifetime) for $\Bz$ mesons is approximately 
$50\mu$m.  Again, this corresponds to a time resolution of about 
one-half a $\Bz$ lifetime.  As estimated by the \superb proponents, 
this is sufficient for measurement of $\sin 2\beta$ to a precision 
of $0.1^{\circ}$~\cite{O'Leary:2010af}.
The $\Dz$ lifetime is, however, 3.8 times 
shorter than that of the $\Bz$.  At the $\psi(3770)$, therefore, 
a boost ($\beta\gamma$) that is  approximately four times larger 
is required to maintain the same time resolution - 
providing the detector performance is comparable.  Decays of 
\Dz mesons to \CP eigenstates have branching fractions about 
an order of magnitude larger than those of $\Bz$ mesons.  If 
the larger boost is achievable at the $\psi(3770)$, measurements of the 
angle $\beta_c$ with a precision similar to that of $\beta$ are 
conceivable.  A detailed simulation is required to fully 
understand this, however we discuss results of a simple simulation 
study in Section~\ref{sec:numerical} focusing on the potential 
for measuring time-dependent asymmetries using the $\Dz \to \Kp\Km$
and $\Dz \to \pi^+\pi^-$ channels.

\section{Flavor Tagging}
\label{sec:ftag}

Flavor tagging is required to synchronize the time $t$ in the case of 
uncorrelated decays and \deltat for correlated mesons.  
Flavor tagging works on the principle of identifying flavor specific
final states that can unambiguously be used to determine the flavor of
a neutral meson decaying into a \CP state of interest.  A flavor tag
has an associated probability that the assignment is incorrect.  This
so-called mis-tag probability is denoted by $\omega$, and the figure of
merit used to discuss how useful a particular process or set of channels
is for flavor tagging is the dilution $D=1-2\omega$.  More generally one
also considers possible differences between the mis-tag probability of 
a particle $\omega$ and that of the anti-particle $\overline{\omega}$,
where $\Delta \omega = \omega - \overline{\omega}$, and the dilution 
factor becomes $D+\Delta \omega = 1 - 2\omega + \Delta\omega$.  

An important consideration is that the effect of a non-zero value for 
$\Delta\omega$ is an
overall shift in \CP asymmetries at all times and is, therefore, functionally similar to 
the effect of a non-zero value for $|\lamf|^2-1$.  Uncertainty in $\Delta\omega$ is,
therefore, strongly correlated with that in $|\lamf|$.  Any variation in this quantity with decay time must also be well understood if measurements of $\lamf$
are to be meaningful.

\subsection{Flavor tagging of un-correlated mesons}

Flavor tagging of un-correlated \Dz mesons can be accomplished by identification of ``slow''
(low momentum) pions from the processes
$D^{*+}\to \D^0 \pi^+$ or CP conjugate process
is $D^{*-}\to \Dzb \pi^-$. Hence if one can identify a sample of events
where neutral $D$ mesons originate from a $D^{*\pm}$, the charge of the 
associated pion can be used to infer if the $\D$ meson is a \Dz or a \Dzb at
the time of decay where $t=0$.  The technical challenge for experiments is in identifying 
the so-called bachelor $\pi^\pm$ from the $D^{*\pm}$ as this has a low momentum
and is therefore more challenging to reconstruct.
At the \B factories $D^{*}$ from $c\overline{c}$ continuum can be cleanly separated from 
$D^\pm$ from $B$ decay by making a momentum cut above the kinematic threshold
imposed by the $B$ mass. At \lhcb the majority of the $D^*$ mesons of interest
for studying \CP asymmetries are secondary particles produced in the
primary decay of a \B meson. In either case $D^*$ tagged
events will have non-trivial mis-tag probabilities 
arising from mis-reconstruction, wrongly associated slow pions, and from background.
A further source of mis-tagging, though small, could come from mixing of a \Dz used
for tagging. 

One can account for mis-tag probabilities by considering the physical decay rates
rather than the theoretical ones.  These are given by
\begin{eqnarray}
\Gamma^{Phys}(t) &=& (1-\omega)\Gamma (t) + \overline{\omega}\, \overline{\Gamma}(t),\\
\overline{\Gamma}^{Phys}(t) &=& \omega\Gamma (t) + (1 - \overline{\omega})\overline{\Gamma}(t),
\end{eqnarray}
where $\Gamma(t)$ and $\overline{\Gamma}(t)$ are from Eqns~(\ref{EQ:p0toffinal}) and (\ref{EQ:p0bartoffinal}).
It is straightforward to compute the physical \CP asymmetry by inserting these results into
Eq.~(\ref{eq:asym}).  Any precision measurement of a \CP asymmetry using 
this method would require detailed control of the systematic uncertainties associated
with $D^*$ flavor tagging.  This provides a limit on the ultimate precision 
attainable for a given measurement.

\subsection{Flavor tagging of correlated mesons}

The set of flavor specific final states of a $D$ meson can be used to unambiguously identify
if a decay into a \CP state of interest is that of a \Dz or a \Dzb. In analogy with the 
methods used for \B decay tagging (for example see~\cite{Aubert:2009yr}), one can use 
a variety of modes for flavor tagging \D mesons.  The advantage of charm over beauty can 
be seen for example in the use of semi-leptonic decays for flavor tagging.  The
decays $D \to K^{(*)-}\ell^+\nu$ account for 11\% of all D decays, and unambiguously 
assign the the flavor: a \Dz decay is associated with a $\ell^+$ in the final state,
and a \Dzb is associated with a $\ell^-$.  The corresponding situation for 
tagging $D$'s from \B decays is more ambiguous since wrong-sign leptons can arise
from decays of $B$'s to $D^{*} \ell \nu$.
In addition, the flavour of each \Dz is unambiguously known at $\deltat=0$ in
the correlated case.  For uncorrelated \Dz's, however, the one decaying to
a \CP eigenstate may have mixed so that its flavour at $t=0$ is unknown.
Thus 11\% of all events recorded 
at the $\psi(3770)$ can be flavor tagged with a mis-tag probability of essentially zero.
Events with kaon or pions in the final state can also be used for flavor tagging,
however for these the mis-tag probability will be non-zero.

From the perspective of performing a precision measurement, which will be an inevitable 
requirement for testing the SM, minimization of systematic uncertainties will be of 
paramount importance.  Here the benefit of accumulating data at charm threshold is clear as one 
can choose to restrict the analysis to using only semi-leptonic tag decays with an 11\% efficiency.  In 
doing so an essentially pure \CP sample can be reconstructed with $\omega \simeq \overline{\omega} \simeq 0$.  

The viability of including other final states in the tagging algorithm, for example 
$\Dz\to K^{*-} (\pi^+, \rho^+)$ etc. introduces experimental issues that may need to be understood.  
These decays can proceed by a tree level Cabibbo allowed transition, and the \CP
conjugate final state can proceed via a doubly Cabibbo suppressed transition.  This 
introduces an ambiguity in the flavor tag assignment (hence dilution), and 
as \D mesons can mix there are several amplitudes from initial to final state.
This raises the issue of possible tag-side interference which is a well known
effect for hadronic \B tagging~\cite{Long:2003wq}.

One can account for mis-tag probabilities by considering the physical decay rates
as a function of \deltat.  These are given by
\begin{eqnarray}
\Gamma^{Phys}(\deltat) &=& (1-\overline{\omega})\Gamma (\deltat) + \omega \, \overline{\Gamma}(\deltat),\label{eq:taggedrates}\\
\overline{\Gamma}^{Phys}(\deltat) &=& \overline{\omega}\, \Gamma (\deltat) + (1 - \omega)\overline{\Gamma}(\deltat),
\label{eq:bartaggedrates}
\end{eqnarray}
where $\Gamma(\deltat)$ and $\overline{\Gamma}(\deltat)$ are from 
Eqns~(\ref{EQ:p0toffinaldeltat}) and (\ref{EQ:p0bartoffinaldeltat}).
Note that the mistag probabilities are interchanged when moving from the uncorrelated (same side tagging)
to the correlated (opposite side tagging) case.
The \CP asymmetry obtained when allowing for tagging dilution is given by

\begin{widetext}
\begin{eqnarray}
{\cal A}^{Phys}(\deltat) &=& \frac{\overline{\Gamma}^{Phys}(\deltat) - \Gamma^{Phys}(\deltat) } { \overline{\Gamma}^{Phys}(\deltat) + \Gamma^{Phys}(\deltat)}, \\
  &=&-\Delta \omega + \frac{ (D + \Delta\omega)e^{\Delta \Gamma \deltat/2}[ (|\lamf|^2 - 1)\cos\Delta M\deltat + 2 Im\lamf \sin\Delta M \deltat ]}{h_+ (1+|\lamf|^2)/2 + Re(\lamf) h_-}.
  \label{eq:asymtagging}
\end{eqnarray}
\end{widetext}

Hence a non-zero mistag probability results in a dilution of the amplitude of oscillation,
and any particle-anti-particle mistag probability difference results in an overall offset in
the asymmetry\footnote{Note that for uncorrelated decays, one interchanges $\omega$ and $\overline{\omega}$,
hence the sign of the $\Delta \omega$ terms changes.}. 
Eq.~(\ref{eq:asymtagging}) highlights the attraction of using data from charm 
threshold to minimize systematic uncertainties associated with tagging.
To a good approximation $\Delta \omega = 0$, and $D=1$ for semi-leptonic tagged decays, 
hence the error on $\lamf$ from this source will be relatively small.  Furthermore as mentioned 
above, there is only a single amplitude contributing to the semi-leptonic tagged side of the 
event, hence tag-side interference is not an issue.  Thus if one observes a non-zero asymmetry,
this can readily be identified as a physical effect. For any other tagging category, 
a significant amount of work would need to be done in order to establish firstly 
if the systematic uncertainties were under control in terms of tagging performance,
and secondly if there is a significant issue related to tag-side interference that 
could otherwise manifest large fake signals of \CP violation.

\section{Analysis of \CP eigenstates}
\label{sec:analysis}

\begin{figure*}[!hb]
\begin{center}
  \resizebox{15.0cm}{!}{
\includegraphics{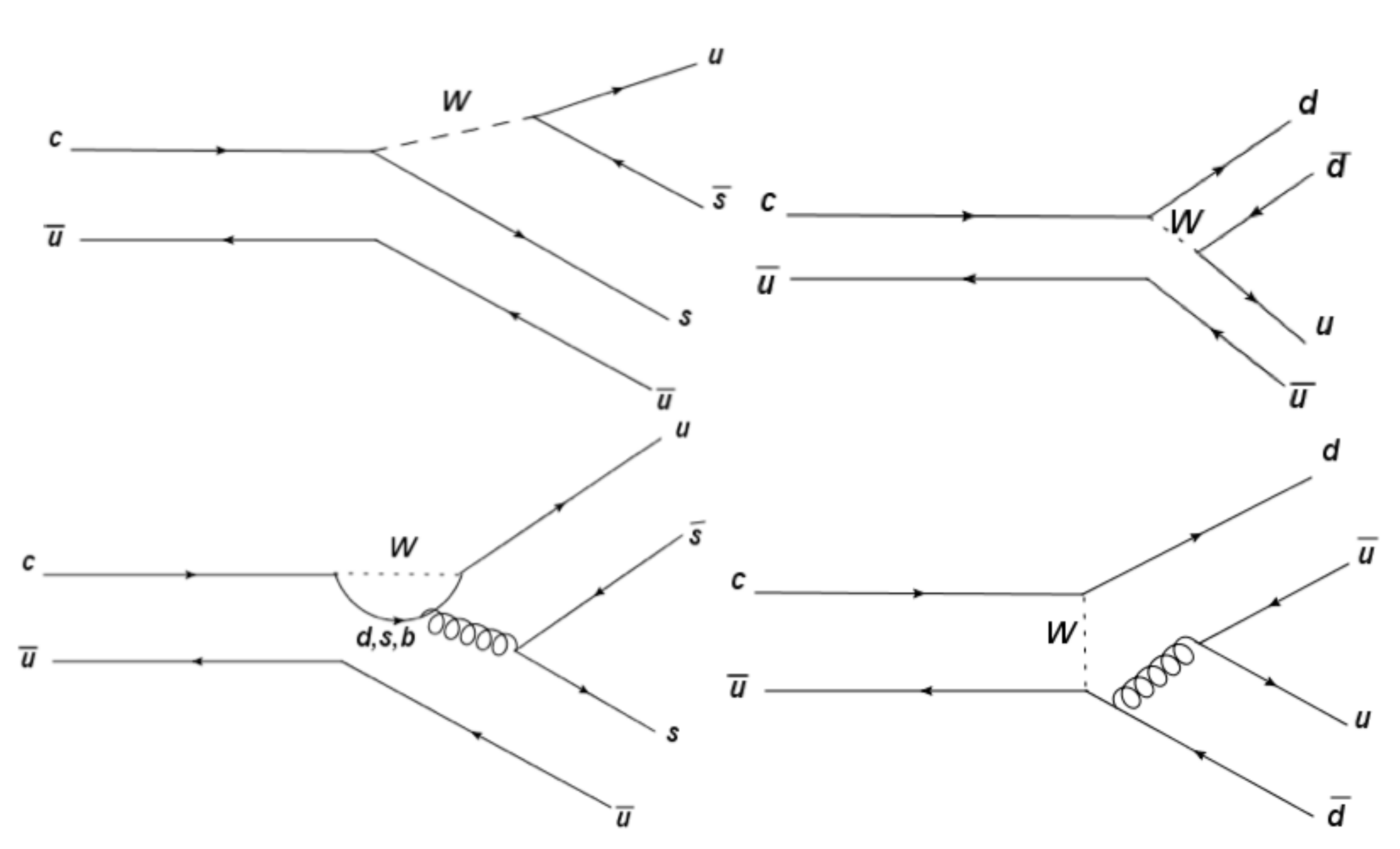}
}
  \caption{Feynman diagrams for (top-left to bottom-right) tree, color suppressed tree, penguin and $W$ exchange topologies.}
  \label{fig:feyn}
\end{center}
\end{figure*}

We have considered a number of two and three body \CP eigenstate decays
of neutral $D$ mesons in order to determine the CKM element contributions to the 
decay amplitude, and hence the corresponding weak phase information that could be
extracted from a given decay.  The full set of modes is listed in 
Table~\ref{tbl:eigenstates}, where we have considered contributions from
tree, color suppressed tree, loop (penguin) and weak-exchange topologies.
Possible long distance contributions have been neglected in this paper.
The Feynman diagrams for these topologies, in the case of two body final
states, are shown in Fig.~\ref{fig:feyn}.

\begin{table*}[!ht]
\caption{\CP eigenstate modes considered in this paper indicating the topologies
contributing to each process in terms of the CKM factors associated with 
$T$ (tree), $CS$ (color suppressed tree), $P_{q}$ (penguin where $q$ is a down-type quark),
and $W_{EX}$ (W-exchange) transitions.  
Blank entries in the table denote that a given topology does not 
contribute to the total amplitude of the decay, and the relative strengths of these amplitudes
decrease from left to right. Non-resonant modes are indicated by NR in order to differentiate
from the resonant contributions with the same final state (but different \CP eigenvalue and 
CKM element contribution).}\label{tbl:eigenstates}
\begin{center}
\renewcommand{\arraystretch}{1.2}
\begin{tabular}{lccccc}
mode & $\eta_{CP}$ & $T$ & $CS$ & $P_{q}$ & $W_{EX}$ \\ \hline \hline
$\Dz\to K^+ K^-$             & $+1$ & $\vcs\vus^*$ &               & $\vcq\vuq^*$  & \\
$\Dz\to \KS \KS$             & $+1$ &              &               &               & $\vcs\vus^* + \vcd\vud^*$  \\
$\Dz\to \pi^+ \pi^-$         & $+1$ & $\vcd\vud^*$ &               & $\vcq\vuq^*$  & $\vcd\vud^*$ \\
$\Dz\to \pi^0 \pi^0$         & $+1$ &              & $\vcd\vud^*$  & $\vcq\vuq^*$  & $\vcd\vud^*$ \\
$\Dz\to \rho^+ \rho^-$       & $\pm 1$ & $\vcd\vud^*$ &               & $\vcq\vuq^*$  & $\vcd\vud^*$ \\
$\Dz\to \rho^0 \rho^0$       & $\pm 1$ &              & $\vcd\vud^*$  & $\vcq\vuq^*$  & $\vcd\vud^*$ \\
$\Dz\to \phi \piz$           & $+1$ &              & $\vcs\vus^*$  & $\vcq\vuq^*$  &  \\
$\Dz\to \phi \rhoz$          & $\pm 1$ &              & $\vcs\vus^*$  & $\vcq\vuq^*$  &  \\
$\Dz\to f^0(980) \piz$       & $-1$ &              & $\vcs\vus^*+\vcd\vud^*$ & $\vcq\vuq^*$ \\
$\Dz\to \rho^0 \piz$         & $+1$ &              & $\vcd\vud^*$  & $\vcq\vuq^*$  & $\vcd\vud^*$ \\
$\Dz\to a^0 \piz$            & $-1$ &              & $\vcd\vud^*$  & $\vcq\vuq^*$  & $\vcd\vud^*$ \\
$\Dz\to \KS\KS\KS$           & $+1$ &              &               &               & $\vcs\vud^* + \vcd\vus^*$ \\
$\Dz\to \KL\KS\KS$           & $-1$ &              &               &               & $\vcs\vud^* + \vcd\vus^*$ \\
$\Dz\to \KL\KL\KS$           & $+1$ &              &               &               & $\vcs\vud^* + \vcd\vus^*$ \\
$\Dz\to \KL\KL\KL$           & $-1$ &              &               &               & $\vcs\vud^* + \vcd\vus^*$ \\
\hline
$\Dz\to \KS \piz$            & $-1$ &              & $\vcs\vud^*+\vcd\vus^*$  &    & $\vcd\vus^*$  \\
$\Dz\to \KS \omega$          & $-1$ &              & $\vcs\vud^*+\vcd\vus^*$  &    & $\vcd\vus^*$  \\
$\Dz\to \KS \eta$            & $-1$ &              & $\vcs\vud^*+\vcd\vus^*$  &    & $\vcs\vud^*+\vcd\vus^*$ \\
$\Dz\to \KS \eta^\prime$     & $-1$ &              & $\vcs\vud^*+\vcd\vus^*$  &    & $\vcs\vud^*+\vcd\vus^*$\\
$\Dz\to \KS \pi^+\pi^-$ (NR) & $+1$ &              & $\vcs\vud^*$             &    & $\vcd\vus^*+\vcs\vud^*$\\
$\Dz\to \KS \rho^0$          & $-1$ &              & $\vcs\vud^*+\vcd\vus^*$  &    & $\vcd\vus^*$   \\
$\Dz\to \KS K^+K^-$ (NR)     & $-1$ & $\vcd\vus^*$ & $\vcs\vud^*$           &    &                \\
$\Dz\to \KS \phi$            & $-1$ &              & & & $\vcs\vud^*+\vcd\vus^*$\\
$\Dz\to \KS f^0$             & $+1$ &              & $\vcd\vus^*$  &               & $\vcd\vus^*+\vcs\vud^*$    \\
$\Dz\to \KS a^0$             & $+1$ &              & $\vcd\vus^*$  &               & $\vcd\vus^*+\vcs\vud^*$    \\
\hline
$\Dz\to \KL \piz$            & $+1$ &              & $\vcs\vud^*+\vcd\vus^*$  &    & $\vcd\vus^*$  \\
$\Dz\to \KL \omega$          & $+1$ &              & $\vcs\vud^*+\vcd\vus^*$  &    & $\vcd\vus^*$  \\
$\Dz\to \KL \eta$            & $+1$ &              & $\vcs\vud^*+\vcd\vus^*$  &    & $\vcs\vud^*+\vcd\vus^*$ \\
$\Dz\to \KL \eta^\prime$     & $+1$ &              & $\vcs\vud^*+\vcd\vus^*$  &    & $\vcs\vud^*+\vcd\vus^*$\\
$\Dz\to \KL \pi^+\pi^-$ (NR) & $-1$ &              & $\vcs\vud^*$             &    & $\vcd\vus^*+\vcs\vud^*$\\
$\Dz\to \KL \rho^0$          & $+1$ &              & $\vcs\vud^*+\vcd\vus^*$  &    & $\vcd\vus^*$   \\
$\Dz\to \KL K^+K^-$ (NR)     & $+1$ & $\vcd\vus^*$ & $\vcs\vud^*$           &    &                \\
$\Dz\to \KL \phi$            & $+1$ &              & & & $\vcs\vud^*+\vcd\vus^*$\\
$\Dz\to \KL f^0$             & $-1$ &              & $\vcd\vus^*$  &               & $\vcd\vus^*+\vcs\vud^*$    \\
$\Dz\to \KL a^0$             & $-1$ &              & $\vcd\vus^*$  &               & $\vcd\vus^*+\vcs\vud^*$    \\
\hline
\end{tabular}
\end{center}
\end{table*}


It is clear 
from Table~\ref{tbl:eigenstates} that the modes
we are considering do not contain contributions from all four topologies,
which simplifies the situation somewhat.  One should note that the $\piz\piz$ final state 
typically consists of four photons, however it would be possible to reconstruct a vertex
and perform a time-dependent analysis for events where photon conversion in detector material had 
occurred.  Also, in about one in 40 instances, one of the \piz's will internally convert in a Dalitz
decay $\piz\to\epem\gamma$, in which the $e$-pair with non-zero opening angle will provide an excellent 
location of the vertex position.

In general we are interested in the value of $\lamf$ as given in Eq.~(\ref{eq:lambda}) when exploring 
\CP violation.  This can be written as
\begin{eqnarray}
\lamf = \left| \frac{q}{p}\right|e^{i\phi_{MIX}} \left| \frac{\overline{A}}{A}\right|e^{i\phi_{CP}},
\end{eqnarray}
where $\phi_{MIX}$ is the phase of $\Dz\Dzb$ mixing, and $\phi_{CP}$ is the overall phase of the 
$\Dz \to f_{CP}$ decay, where $f_{CP}$ is a \CP eigenstate.  The amplitude $A$ in general can have
contributions from different topologies, and as a result $\phi_{CP}$ is not necessarily directly 
related to an angle of the charm unitarity triangle.  This can be seen from the following
\begin{eqnarray}
A &=& |T|e^{i\phi_T} + |CS|e^{i\phi_{CS}} + |W|e^{i\phi_{W}} \nonumber \\
  && + \sum\limits_{q=d, s, b} |P_q|e^{i\phi_{q}},\label{eq:amplitude}
\end{eqnarray}
where the $\phi_j$, $j = T, CS, W, q$ are phases of the tree, color suppressed tree, $W$ exchange
and penguin amplitudes respectively, and the coefficients of the exponentials are the magnitudes
corresponding to those amplitudes.  In general one should note that $\phi_j$ consists of a 
strong phase ($\delta_j$ which is invariant under \CP) and a weak phase ($\phi_j^W$ which 
changes sign under \CP), thus $\phi_j = \phi_j^W + \delta_j$. 

If one considers the tree dominated decays such as $D\to K^+K^-$, $\pi^+\pi^-$, $\Kz K^+K^-$, and $\Kz \pi^+\pi^-$,
assuming that there is a negligible penguin or color suppressed tree (and in the case of a $\pi^+\pi^-$ final state, 
one also neglects $W$ exchange) contribution then it follows that
\begin{eqnarray}
\lamf = \left| \frac{q}{p}\right|e^{i\phi_{MIX}} e^{-2i\phi_{T}^W},
\end{eqnarray}
where $|T|$ and the strong phase $e^{i\delta_T}$ cancel in the ratio of $\overline{A}/A$.
While this may be adequate for a rudimentary \CP asymmetry measurement, eventually
it would be necessary to understand the role of the penguin contribution to the two body final states,
and that of the color suppressed tree for the three body non-resonant case.
It is also clear that in order to interpret any \CP asymmetry measurement 
in terms of an angle of the charm triangle, one needs to obtain a precision measurement
of $q/p$ in the neutral charm meson system.
It should be noted that the same arguments also apply for excited states where for example
pseudoscalar mesons are replaced by vector or axial-vector particles.
For final states with two spin one particles one must perform an angular 
analysis in order to disentangle \CP even and \CP odd components of the decay.

In the more general case of two amplitudes contributing to the final state (here we consider the case for a 
tree and a single penguin contribution $P$ as a simplification), then
\begin{eqnarray}
\lamf &=& \left| \frac{q}{p}\right|e^{i\phi_{MIX}} \frac{e^{-i\phi_T} + r e^{-i\phi_P} }{ e^{i\phi_T} + r e^{i\phi_P}},
\end{eqnarray}
where the penguin to tree ratio $r=|P|/|T|$ is an unknown quantity that needs to be evaluated from data.

If we now return to the amplitudes in Table~\ref{tbl:eigenstates}, it is possible to 
determine the relative strengths of the different contributions by considering the number
of vertices in the corresponding Feynman diagram and the CKM factors related to these vertices.
The distinct products of CKM factors appearing in the table are summarized, up to ${\cal O}(\lambda^6)$,
in the following
\begin{eqnarray}
 \vcs\vus^* &=& \lambda - \frac{\lambda^3}{2} - \left(\frac{1}{8} +\frac{A^2}{2}\right) \lambda^5,\\
 \vcd\vud^* &=& -\lambda + \frac{\lambda^3}{2} + \frac{\lambda^5}{8} + \frac{A^2\lambda^5}{2}[1 -2(\bar\rho +i\bar\eta)],\\
 \vcb\vub^* &=& A^2 \lambda^5(\bar\rho + i\bar\eta),\\
 \vcs\vud^* &=& 1 - \lambda^2- \frac{A^2\lambda^4}{2} \nonumber \\
           && + A^2\lambda^6\left[\frac{1}{2} - \bar \rho - i\bar \eta -\bar\eta^2 - \bar\rho^2 \right],\\
 \vcd\vus^* &=& -\lambda^2 + \frac{A^2\lambda^6}{2} [1 -2(\bar\rho +i\bar\eta)].
\end{eqnarray}
Four of the five amplitudes are complex; $\vcb\vub^*$ has a large phase ($\gamma_c$),
while $\vcd\vud^*$ and $\vcd\vus^*$ (both have the phase
of \vcd\ which is $\beta_c-\pi$) are related to a small weak phase.  The remaining term
$\vcs\vud^*$ also has a small phase, entering at ${\cal O}(\lambda^6)$ in the 
amplitude.
It is interesting to note that the amplitude $\vcb\vub^*$ only proceeds via a
penguin process, and is always accompanied by a tree (color allowed or suppressed) 
and two other penguin amplitudes which will dominate.  Hence it is unlikely
that one will ever be able to collect data with sufficient statistical precision to measure $\gamma_c$ 
from processes involving a $c\to u$ penguin transition.

The next most promising phase to measure is associated with transitions 
mediated by $\vcd\vud^*$ where the imaginary component of this 
amplitude is ${\cal O}(\lambda^5)$.  Modes involving this transition at leading
order include $D\to \pi^+\pi^-$, $\rho^+\rho^-$, $h^0h^0$, where $h = \pi^0, \rho^0, a^0$.
These are discussed in Sections~\ref{sec:dzerotopipi} 
and ~\ref{sec:dzerotorhorho}.

The combination of CKM elements with the smallest phase to this order in $\lambda$
is $\vcd\vus^*$ which is doubly Cabibbo suppressed.  This CKM factor appears in 
the $W$ exchange amplitudes for $\Dz\to 3\Kz$, however it does so in conjunction 
with other exchange amplitudes that are Cabibbo allowed.  This story
is repeated for almost all of the other $\Dz$ modes we consider with a neutral or 
charged kaon in the final state.  The exceptions $\Dz\to \Kz f^0$ and $\Kz a^0$
have a color suppressed tree proceeding with a CKM factor of $\vcd\vus^*$, 
and a $W$ exchange amplitude with a factor of both $\vcd\vus^*$ and $\vcs\vud^*$.
Hence while the $\Delta S \neq 0$ modes contain weak phase information, it 
will be difficult to experimentally distinguish between the amplitudes
contributing to the decay and extract a precision measurement of $\beta_c$.

\subsection{$\Dz\to K^+K^-$ and related modes}
\label{sec:dzerotokk}

$\Dz\to K^+K^-$ measures the phase of $\vcd\vud^*$ only in a sub-dominant penguin transition, and 
is otherwise dominated by a real tree amplitude with a CKM factor of $\vcs\vus^*$.  
Hence to first order one would expect to observe an asymmetry consistent with the mixing
phase $\phi_{MIX}$, with no CKM weak phase contribution.  This channel provides, therefore, 
a useful cross check of detector reconstruction and calibration.  It also provides 
measurements of $|q/p|$ and $\phi_{MIX}$ to complement others that may be available.
Given that the SM prediction of the asymmetry in this channel is small, this
is also an ideal mode to use when searching for NP.
It is interesting to note that \vcs\ is complex at ${\cal O}(\lambda^6)$
 using the convention of~\cite{Buras:1994ec}.  Ultimately a measurement 
of $\beta_c$ could be possible, however this is not likely to be the most promising mode to measure 
the angle.

The same is true for the vector-vector final state $K^{*+}K^{*-}$.
Using the naive factorization framework the fraction of longitudinally polarized
events $f_L$ in the decay of a spin zero meson decaying into two vector mesons can be
estimated as~\cite{Suzuki:2002yk}
\begin{eqnarray}
f_L &=& 1 - \frac{m_V^2}{M^2},\label{eq:polarisation}
\end{eqnarray}
where $m_V$ is the vector meson mass, and $M$ is the mass of the decaying parent particle.
Using this we can estimate $f_L$ for $\Dz\to K^{*+}K^{*-}$ to be $\sim 0.77$.
Hence one would be required to perform an angular analysis in order to extract \CP asymmetry 
parameters from this decay.

\subsection{$\Dz\to \pi^+\pi^-$ and related modes}
\label{sec:dzerotopipi}

$\Dz\to \pi^+\pi^-$ measures the phase of $\vcd\vud^*$ in the leading order tree, one of the penguin
amplitudes, and the $W$ exchange topologies.  Of the remaining two penguin amplitudes
that contribute to this decay, one is completely negligible (mediated by a $b$ quark
loop) and the other is of the order of $\lambda$.  The non-trivial penguin 
topologies are doubly Cabibbo suppressed loops and proceed at order $\lambda^2$,
where as the tree amplitude is singly Cabibbo suppressed.  A rudimentary measurement
of this process could in principle ignore the penguin contribution, in which case
$Im\lamf \simeq \sin(\phi_{MIX} - 2\beta_c)$.  Thus there will be a four-fold 
ambiguity in any measurement of $\beta_c$. However one should note that a 
more complete analysis would be required in order to extract the weak
phase and disentangle the contribution from the $c\to s \to u$ penguin.

Bigi and Sanda have pointed out~\cite{Bigi:1999hr} that there are two Isospin amplitude contributions
to $D\to \pi^+\pi^-$.  Actually the situation is almost exactly the same as the $B\to \pi\pi$, as we have
an Isospin $1/2$ meson (a $B$ or a $D$) decaying into two pions. 
The only differences are that, in general, we need to assume $\Delta \Gamma \neq 0$, 
for charm decays, which is a generalization that the existing measurements of $\Bz\to \pi\pi$ have not yet considered,
and we neglect the $W$ exchange amplitude (which has the same weak phase as the tree).  
The ramification of this is straightforward $-$ instead of 
measuring $S$ and $C$ of Eq.~(\ref{eq:bzcpparameters}) in order to determine the weak phase,
one measures the real and imaginary parts of $\lambda$ as given in Eq.~(\ref{eq:lambda}).
One also measures the amplitudes for the Isospin related $\pi^+\pi^-$, $\pi^+\pi^0$, $\pi^0\pi^0$ 
decays to perform an Isospin amplitude decomposition of $\pi\pi$ 
final states, as described below,
in order to disentangle the phase contribution from the tree and penguin amplitudes.

Similar considerations apply to other final states with 2-body combinations 
of $\pi^\pm$, $\rho^\pm$, and $a_1^\pm(1260)$.  Such states that include 
two spin one particles would require an angular analysis in order to 
disentangle \CP even and odd parts and correctly measure the time-dependent
\CP asymmetry parameters.  For example in the $\D\to \rho\rho$ case,
we expect $f_L \sim 0.83$.
As in the $B$ meson system, one can apply the same Isospin analysis procedure
in order to bound penguins for $\D\to\rho\rho$ decays, although one should
take care to establish whether there is evidence of any $I=0$ component arising from
the finite width of the $\rho$~\cite{falk}.  Based on the penguin hierarchy observed
in $B$ decays, we expect that, unless long
distance effects play an important role in $c\to u\overline{u}d$ transitions,
that $\D\to\rho\rho$ might have a smaller penguin contribution than $\D\to\pi\pi$.
If this turns out to be the case, then $\D\to \rho^+\rho^-$ may provide a more
precise constraint on $\beta_c$ than $D\to \pi^+\pi^-$, and should not
be overlooked by experimentalists.  It should be noted that, while a Quasi-2-Body 
approach (where the intermediate resonances are treated as particles) may be
sufficient for a preliminary study, a full amplitude analysis would eventually be 
required in order to extract weak phase information from $\D\to\rho\rho$ decays.  

For decays like, for example, $D\to\rho\pi$, the isospin structure can be more 
complex, in general 
\cite{Lipkin:1991st}.
We note, however, that a complete decay amplitude analysis of the $\piz\pip\pim$ 
Dalitz plot has been performed by both CLEO 
\cite{CroninHennessy:2005sy}
and by \babar 
\cite{Aubert:2007ii}
and that, in a subsequent isospin analysis of this 3-body final state
\cite{Gaspero:2008rs,Gaspero:2010pz},
it has been found that the amplitude is dominated by a single ($I=0$) component.  This 
situation is found to be consistent with a decay model with no penguin contribution
\cite{Bhattacharya:2010id}
but by T, W and CS amplitudes, all with the same phase.  This makes this channel
particularly suitable for extraction of $\beta_c$.  The \babar $\Upsilon(4S)$ sample 
was very clean and a factor five larger than for the $\pip\pim$ channel.  A similar 
statement can be made for CLEOc running at charm threshold.  
For LHCb the trigger is known to be less efficient for multi-body final
states, which in general produce fewer tracks with high transverse 
momenta to trigger on.  As a result we do not expect LHCb to be able to make a
competitive measurement of $D\to \pi^+\pi^-\piz$ decays when compared with the
potential of future \epem\ experiments. The analysis of this channel is 
certainly more complex than that for $\pip\pim$, but it has been found in both 
\babar and in \belle experiments that the multi-body channels add useful constraints
and provide reliable results.

\subsubsection{An Isospin analysis of $\D\to \pi\pi$ and $\D\to \rho\rho$ decays}
\label{sec:isospinanalysis}

For these decays, the Tree and Penguin decay amplitudes are distinguished by their isospin changing structures.
The prescription given here parallels the
one described in Ref.~\cite{gronaulondon} which outlines
how to measure the unitarity triangle angle $\alpha$ from $\B\to \pi\pi$ decays and to constrain so-called penguin pollution. 
Bose symmetry dictates that, for either \Bz or \Dz decays the two-pion final states can be in either an $I=0$ or 
an $I=2$ final state.  
In this case, triangular relationships between amplitudes $A^{ij}(\overline{A}^{ij})$ 
for $D(\bar D)\to h^ih^j$ decays ($h=\pi$ or $\rho$) exist:
\begin{eqnarray}
\frac{1}{\sqrt{2}}A^{+-} = A^{+0} - A^{00},\\
\frac{1}{\sqrt{2}}\overline{A}^{-+} = \overline{A}^{-0} - \overline{A}^{00},
\end{eqnarray}
where the charges are $i, j= +1, -1, 0$.  These two triangles can be aligned with a common
base given by $A^{+0}=\overline{A}^{-0}$, in which case the angle between $A^{+-}$ and 
$\overline{A}^{-+}$ is the shift in the measured phase resulting from penguin contributions.

Obviously, one must measure rates for $D^0\to h^+h^-$, $D^+\to h^+h^0$,
and $D^0\to h^0h^0$ in order to extract the weak phase of interest: $\beta_c$.
The amplitude of sinusoidal oscillation given in Eq.~(\ref{eq:asym})
or (\ref{eq:asymdeltat})
is related to $\lamf = \sin(\phi_{MIX}-2\beta_{c,eff})$.  The 
proposed Isospin analysis would enable one to translate a measurement
of $\beta_{c,eff}$ to a constraint on $\beta_{c}$, given a precise determination
of the mixing phase and the amplitudes of $D$ decays to $hh$ final states.
As final states with more than one neutral particle are required for the Isospin
analysis, it will only be possible to measure the weak phase
using $\Dz\to hh$ decays in an $e^+e^-$ environment.  Ultimately the viability
of this method will depend upon theoretical control of any relevant
topologies that have been neglected, for instance long-distance 
and isospin-breaking effects.

\subsection{$\Dz\to \rho^0\rho^0$ and related modes}
\label{sec:dzerotorhorho}

$\Dz\to \rho^0\rho^0$ measures the phase of $\vcd\vud^*$ via the color suppressed tree, one
penguin, and $W$ exchange amplitudes. Of the remaining two penguin amplitudes
that contribute to this decay, one is completely negligible (mediated by a $b$ quark
loop) and the other is of the order of $\lambda$.  Hence the method to extract
the weak phase from this decay is a repeat of the situation for $\Dz\to \pi^+\pi^-$
discussed in Section~\ref{sec:dzerotopipi}.  In order to disentangle the 
penguin contribution to the time-dependent \CP asymmetry measurement, one 
would have to measure $\Dz\to \rho^+\rho^-$, which includes two \piz mesons
in the final state.  So once again, this process can only be used to 
precisely constrain the weak phase in an $e^+e^-$ environment.  It should be noted
that with $\rho^0\rho^0$, one can 
easily measure the time-dependent asymmetry, and use the result 
to reduce the number of ambiguities in the $D\to\rho\rho$ Isospin analysis.

\subsection{New physics}
\label{sec:newphysics}

The topologies summarized in Table~\ref{tbl:eigenstates} are conveniently categorized in a 
way where one can envisage different types of NP affecting the amplitudes 
contributing to the decay rate.   NP can manifest itself in any of the topologies, and
while one normally ignores the possibility of NP in tree contributions it is worth noting that 
the measurement of $\sin2\beta$ from $\B\to J/\psi K^0$ are currently inconsistent
with SM expectations at a level of $3.2\sigma$~\cite{Lunghi:2011xy}.  
This highlights the importance of embarking on a quest to measure both the mixing
phase and $\beta_c$ as proposed here.
In particular the penguin amplitudes could be 
affected by NP in loop transitions mediated via SUSY partners replacing the SM 
quarks and $W^\pm$.  Hence the modes $\Dz \to h^0h^0$, where $h=\piz, \rhoz, \phi$ are 
particularly good candidates to probe NP manifest through this mechanism.  
The remaining modes considered here could be used to detect NP contributions from
amplitudes that compete with the SM tree or exchange amplitudes. In 
general any large observation of \CP violation in charm decays is 
expected to be a sign of NP~\cite{Pais:1975qs}.  If one does observe a signal, then care must
be taken in order to disentangle the weak phase of interest from the 
$\Dz-\Dzb$ mixing phase.  This, in turn, will require significantly better measurement
of mixing parameters than are currently available.  

\section{Constraining the sides of the charm triangle}
\label{sec:sides}

The charm unitarity triangle given in Eq.~(\ref{eq:charmtriangle}) can also be constrained by 
measurements of the sides, essentially magnitudes of the elements of the CKM matrix.  
The difference in the lengths of the two long sides $\vud^*\vcd$ and $\vus^*\vcs$
must be able to accommodate the geometry of the third side indicated in Fig.~\ref{fig:triangles}.
While
the direct measurement of \CP violating effects is the focus of this paper, the indirect measurements
required to constrain the shape of the triangle independently of \CP asymmetry measurements
are also important and worthy of a mention.  We briefly examine each of these elements in turn in
the following to highlight how one can increase current knowledge of the triangle via
indirect measurements.

\begin{description}
 \item {\boldmath{\Vud}}: This has been precisely measured using nuclear beta decay, and the experimental level of 
    precision reached is at the level of $0.022\%$~\cite{pdg}.

 \item {\boldmath{\Vus}}: This quantity can be measured precisely in kaon decays, however that 
   has reached a natural conclusion of being dominated by systematic uncertainties.  The level 
   of precision reached for this quantity by averaging results from kaon and $\tau$ decays is $1\%$~\cite{pdg}. 
   Future precision measurements of \Vus\ may be possible via studies of $\tau$ decays into final states 
   with charged kaons.  Thus the \superb and \belletwo experiments
   will be required to improve our knowledge of this quantity.

 \item {\boldmath{\Vub}}: The limiting factor for improving constraints on this element comes
   from a combination of theoretical and experimental issues relating to $B$ decays into semi-leptonic
   final states related to $b\to u$ transitions.  While there has been a lot of work in this area,
   there is still a lot of room for improvement both in terms of theoretical and experimental developments.
   The current level of uncertainty obtained for this quantity is $11\%$~\cite{pdg}.
   From the experimental perspective the inclusive and exclusive results obtained for \Vub\ are
   not in good agreement with each other~\cite{vera}.  Thus the \superb and \belletwo experiments
   will be required to improve our knowledge of this quantity.

 \item {\boldmath{\Vcd}}: Precision measurements of semi-leptonic $D$ decays can improve our knowledge of 
   \Vcd\ beyond the current level of precision ($4.8\%$~\cite{pdg}).  This measurement can 
   be improved upon by the BES III experiment at IHEP, and also by the 
   \superb and \belletwo experiments.  \superb will have the advantage of being able to 
   accumulate at data sample fifty times larger than BES III at charm threshold.
   It is unlikely that \belletwo would ultimately be competitive with a measurement of
   \Vcd as that experiment has no plans to run at charm threshold.

 \item {\boldmath{\Vcs}}: The most precise determinations of \Vcs\ come from measurements of 
    semi-leptonic $D_s$ decays.  The current level of precision obtained for \Vcs\ is $3.5\%$.
    This can be improved by the BES III experiment at IHEP, and also by the 
   \superb experiment, using data collected just above charm threshold.

 \item {\boldmath{\Vcb}}: The limiting factor for improving constraints on this element comes
   from a combination of theoretical and experimental issues relating to $B$ decays into semi-leptonic
   final states related to $b\to c$ transitions.  While there has been a lot of work in this area,
   there is still a lot of room for improvement both in terms of theoretical and experimental developments.
   The current level of precision achieved by measurements of \Vcb\ is $3.2\%$~\cite{pdg}.
   From the experimental perspective the inclusive and exclusive results obtained for \Vcb\ are
   not in good agreement with each other~\cite{vera}.  Thus \superb and \belletwo
   will be required to improve our knowledge of this quantity.
\end{description}

\Vud is the most precisely constrained quantity required to reconstruct the triangle
using the sides having been measured to $0.022\%$.  Hence improved measurements of this
quantity will not play an important role in improving our understanding of the charm triangle.
All of the other quantities are known to precisions of the order of $1-10\%$.
Thus in order to improve indirect constraints of the charm triangle, 
(i) we need to wait for the \superb and \belletwo experiments to improve the limiting 
factors in terms of measuring the above quantities, and (ii) the corresponding 
theoretical developments should also be pursued in order for experiment to remain a 
limiting factor.  It should also be noted that the BES III experiment will be able to 
improve the precision of measurements of \Vcd and \Vcs from semi-leptonic $D$ and $D_s$ 
decays before the SFF's start collecting data.

Interestingly enough the quantities \Vub\ and \Vcb\ also currently limit the precision of 
the sides constraint of the unitarity triangle for $B$ decays, and again the only
routes to experimental improvements on that test are via \superb and \belletwo.

\section{Numerical analysis}
\label{sec:numerical}

In this section we compare the three experimental scenarios,
(i) charm threshold (ii) the SFF's at \FourS, and (iii) \lhcb,
relating to the measurement of \CP violation in $\Dz\to f_{\CP}$ decays, 
where $f_{\CP}$ is a \CP eigenstate.  We neglect resolution effects related 
to the reconstruction of vertices in the detector and translation of this 
spatial distance into values of $\deltat$ or $t$.  Finally, based on the 
expectations from these simulations, we discuss the direct constraint on the 
apex of the $cu$ triangle in Section~\ref{sec:numerical:triangle}.

For the numerical analysis and the extrapolation to the expected precision in
$\beta_{c,eff}$, we generate a set of one hundred Monte Carlo data samples
in each experimental scenario,  For \superb running
at charm threshold we do this for both semi-leptonic and also kaon decays as tags.  
In each sample, we generate \Dz and \Dzb events with no time-integrated 
asymmetry, each according to their respective time dependences described 
in Sec. III.  We simulate effects of mis-tagging either \Dz or \Dzb, then 
perform a binned fit to the resulting asymmetry given in Eq.~(\ref{eq:asymtagging})
and the corresponding form in terms of $t$.  In these fits, $\arg(\lambda_f)$ 
and $|\lambda_{f}|$ are allowed to vary and the values for $\omega$ 
and $\Delta\omega$ are fixed at those used in the event generation.  We 
repeat this analysis for different possible values of the phase 
$\arg(\lambda_{f})$ from $-10^{\circ}$ to $+10^{\circ}$ in $10^{\circ}$ steps.
As a figure of merit for each experimental scenario, we take the average uncertainty, $\sigma_{\phi}$, in this phase from the 100 fits, observing that this is consistent with the spread of central values from the individual fits.

\subsection{Charm threshold}

$D$ meson pairs produced at the $\psi(3770)$ are quantum-correlated, so that the time evolution 
is given by Eqns~(\ref{EQ:p0toffinaldeltat}) and (\ref{EQ:p0bartoffinaldeltat}).  If one accounts
for tagging dilution, then the time-dependent \CP asymmetry is given by Eq.~(\ref{eq:asymtagging}).  
On restricting time-dependent analyses to the use only of semi-leptonic tagged decays, the asymmetry 
simplifies as there is no dilution, since both $\omega$ and $\Delta \omega$ terms can be 
neglected, and any systematic uncertainty in the asymmetry arising from $D\simeq 1$ becomes small.  
Furthermore the $\epem\to \psi(3770)$ environment
is extremely clean, so that systematic uncertainties from background contributions is also small
and under control.  These are important points to stress as we know that the 
$\CP$ phase of interest is expected to be small, hence in order to make a precision measurement the systematic uncertainties must be minimized.

\begin{figure}[!ht]
\begin{center}
  \resizebox{8cm}{!}{
\includegraphics{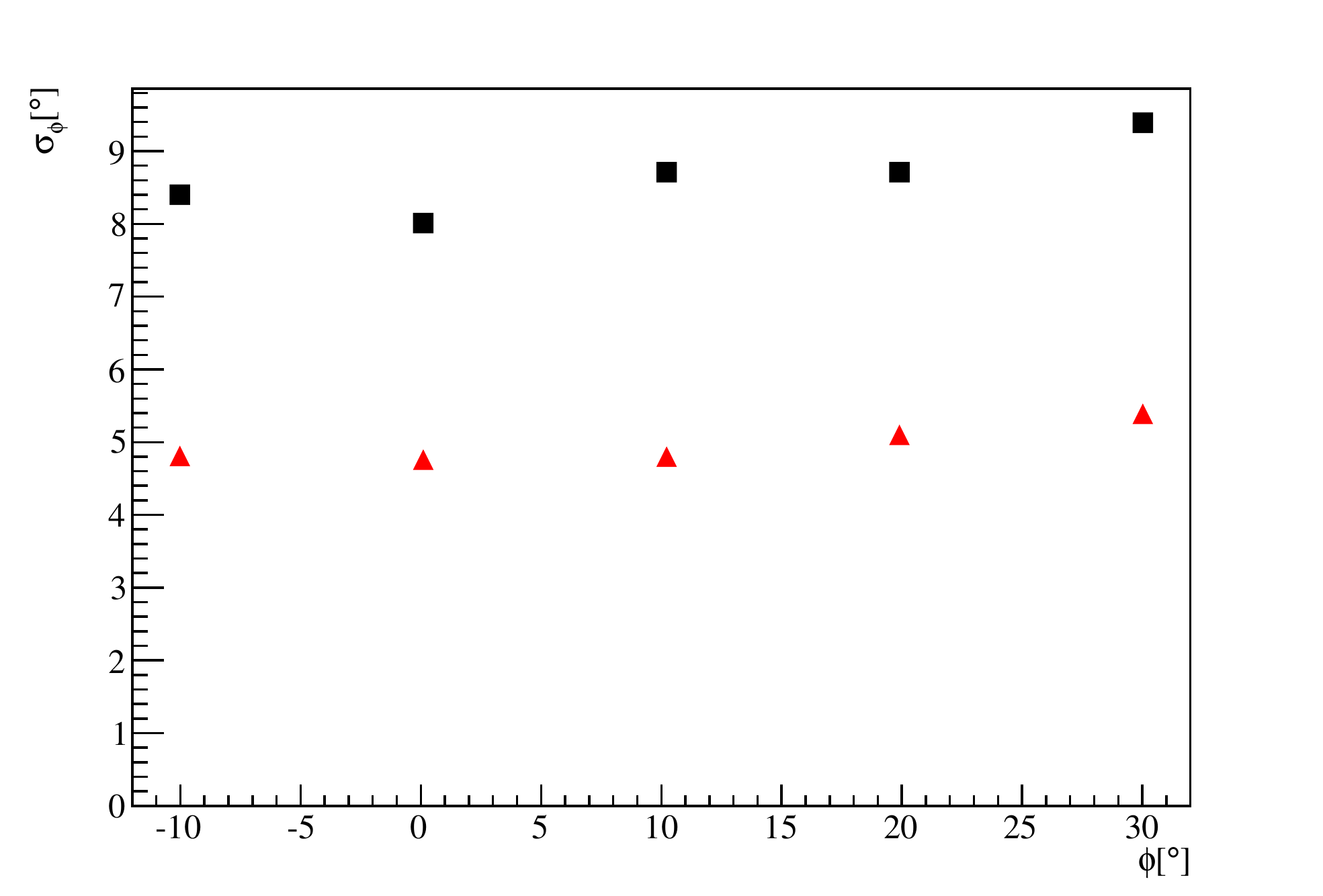}
}
  \caption{The uncertainty in the measured phase $\phi = \phi_{MIX} - 2\beta_c$ as a function of
    the value of $\phi$ for (squares) $D^0\to \pi^+\pi^-$ (triangles) $D^0\to K^+K^-$ decays
    at charm threshold with 500\invfb of data, assuming that the mis-tag probability is negligible,
    and only using the full set of semi-leptonic tagged decays.}
  \label{fig:charmthresholdprecision}
\end{center}
\end{figure}

With 500\invfb of data at charm threshold one can expect to accumulate approximately $1.8\e{9}$
$D$ meson pairs.  With a data sample of 281\invpb CLEO-c obtain 89 $D^0\to \pi^+\pi^-$ candidates with the other $D$ meson decaying semi-leptonically into $X^+ e \nu_e$.  Their efficiency for such events is 50\%~\cite{Asner:2008ft}.  Assuming the same efficiency applies\footnote{Preliminary studies indicate that this is a reasonable assumption.}, we anticipate that \superb could record $158,000$ $X e \nu_e$ tagged $D^0\to \pi^+\pi^-$ events,
corresponding to 489500 events when using the full set of $K^{(*)} \ell \nu_\ell$ tagged events, $\ell = e, \mu$.
We expect about three times the number of events for $D^0\to K^+K^-$.
Figure~\ref{fig:charmthresholdprecision} shows the results obtained for the average uncertainties in the phase $\arg(\lamf) \equiv \phi = \phi_{MIX}+\phi_{CP}$ as a function of that phase.

These results are only for semi-leptonic tags.  We also consider use of hadronically tagged events, for example $\Dz\to K^- X$ ($K^+X$), where $X$ is anything, which correspond to 54\% (3\%) of all neutral \D meson decays.  From
these modes alone, one would expect $\omega \simeq 0.03$, and that the asymmetry in
particle identification of $K^+$ and $K^-$ in the detector will naturally lead to 
a small, but non-zero value of $\Delta \omega$.  We expect that there would be 
approximately 2.2 million kaon tagged $\Dz\to \pi^+\pi^-$ events in 500\invfb 
at charm threshold.  Using these data alone, one would be able to measure
$\phi$ to a precision of $4^\circ$.  Hence, if one combines the results from
semi-leptonic and kaon tagged events, a precision of $\sigma_\phi\sim 3.4^\circ$
is achievable.
This represents a significant improvement in precision over just using 
semi-leptonic tagged events\footnote{Use of the $K$ tag events will introduce tag side interference.  For
$B_d$ analyses this amounts to a few parts per mille, but it will need to be
evaluated for the specific $D$ modes that are used.}.

\subsection{Uncorrelated decays at the \FourS}

\begin{figure}[!ht]
\begin{center}
  \resizebox{8cm}{!}{
\includegraphics{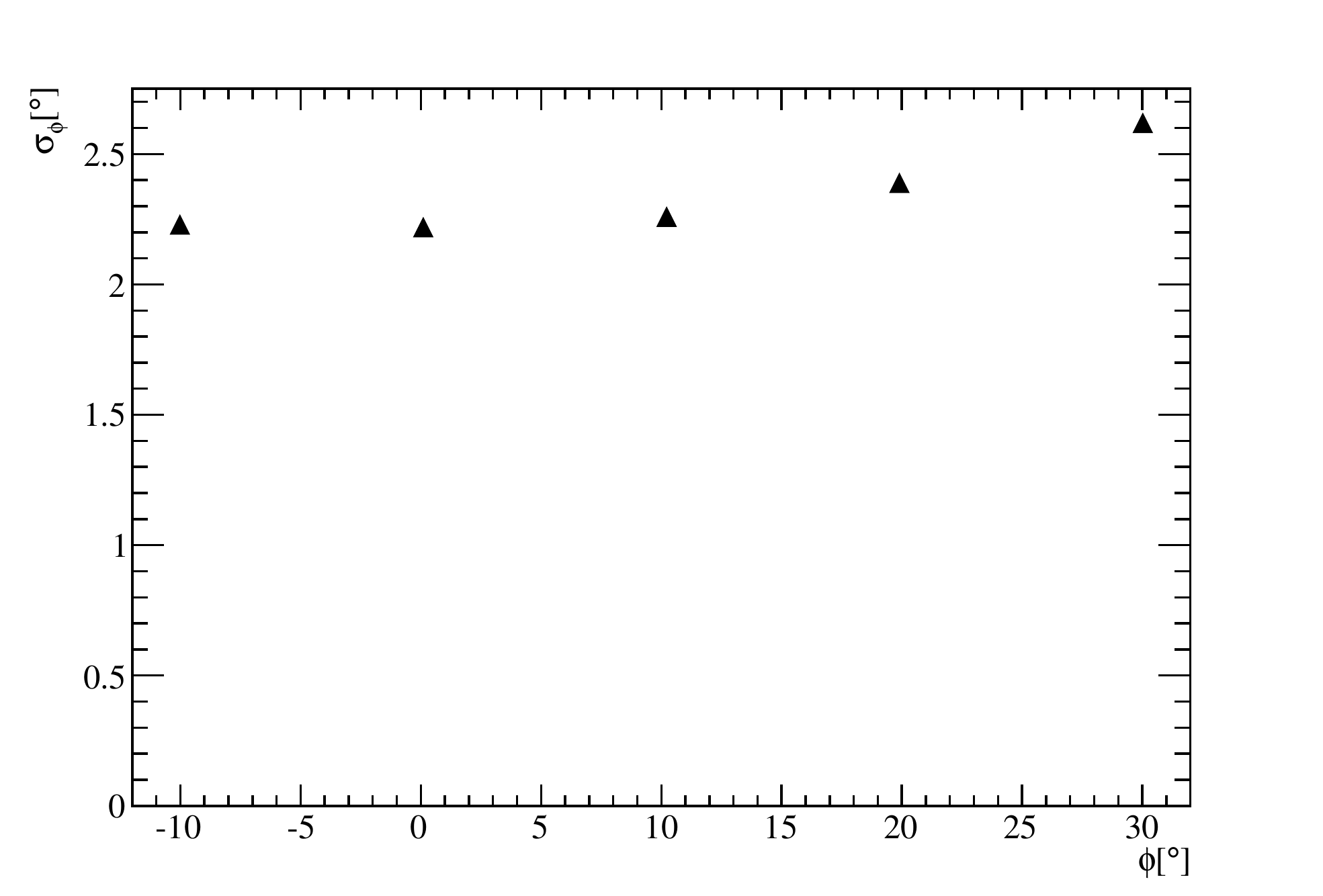}
}
  \caption{The uncertainty in the measured phase $\phi = \phi_{MIX} - 2\beta_c$ as a function of
    the value of $\phi$ for $D^0\to \pi^+\pi^-$ decays
    at the \FourS with 75\invab of data assuming $\omega = \overline{\omega} = 0.01$.}
  \label{fig:foursprecision}
\end{center}
\end{figure}

The scenario at the \FourS is somewhat more complicated than
the situation encountered at the $\psi(3770)$.  Firstly, in order to remove background from
$D$ mesons produced in $B$ meson decay, one restricts the analysis to mesons with high momentum.
In addition to non-trivial backgrounds, one also has to consider non-zero tagging dilution, where
the asymmetry is similar to that given in Eq.~(\ref{eq:asymtagging}), but with $t$ substituted for $\deltat$, and $\omega$
and $\overline{\omega}$ interchanged (hence a sign flip for the $\Delta \omega$ terms).
Thus it is not obvious that $\Delta \omega$ can be neglected, and indeed $D \neq 1$.  
\babar recorded 30,679 $D^*$ tagged $D^0\to \pi^+\pi^-$ events at the \FourS in 384\invfb of data~\cite{Aubert:2007en},
with a purity of 98\%, and where the mis-tag probability for these events is $\sim 1\%$~\cite{delAmoSanchez:2010xz}. 
From this we estimate that one could reconstruct $6.6\e{6}$ $D^*$ tagged $D^0\to \pi^+\pi^-$ events
in a data sample of 75\invab.   
We obtain the sensitivities for $\arg(\lamf)\equiv\phi$ as a function of the phase
shown in Fig.~\ref{fig:foursprecision} assuming this yield and dilution.

To offset the aforementioned background and dilution issues, the increased boost in this \FourS scenario does slightly reduce the
effects of time resolution that are ignored in our analysis here.

\subsection{Uncorrelated decays at \lhcb}

The final scenario considered is that of measuring time-dependent asymmetries from uncorrelated
$D$ mesons in a hadronic environment.  Preliminary time-integrated results from CDF~\cite{CDF}
and \lhcb~\cite{LHCb} indicate that such a measurement is possible.  Dilution and background effects 
will, however, be more severe in this hadronic environment than at an \epem machine.  The measurement of $|\lamf|$ is 
expected to be dominated by such systematic uncertainties, though 
$\arg(\lamf)$ may be less affected, provided that any variation of $\omega$
or $\Delta\omega$ as a function of decay time can be carefully controlled.  It is not clear at this point what the ultimate precision obtained from \lhcb will be.  The best way to ascertain this would be to perform the measurement.

\begin{figure}[!ht]
\begin{center}
  \resizebox{8cm}{!}{
\includegraphics{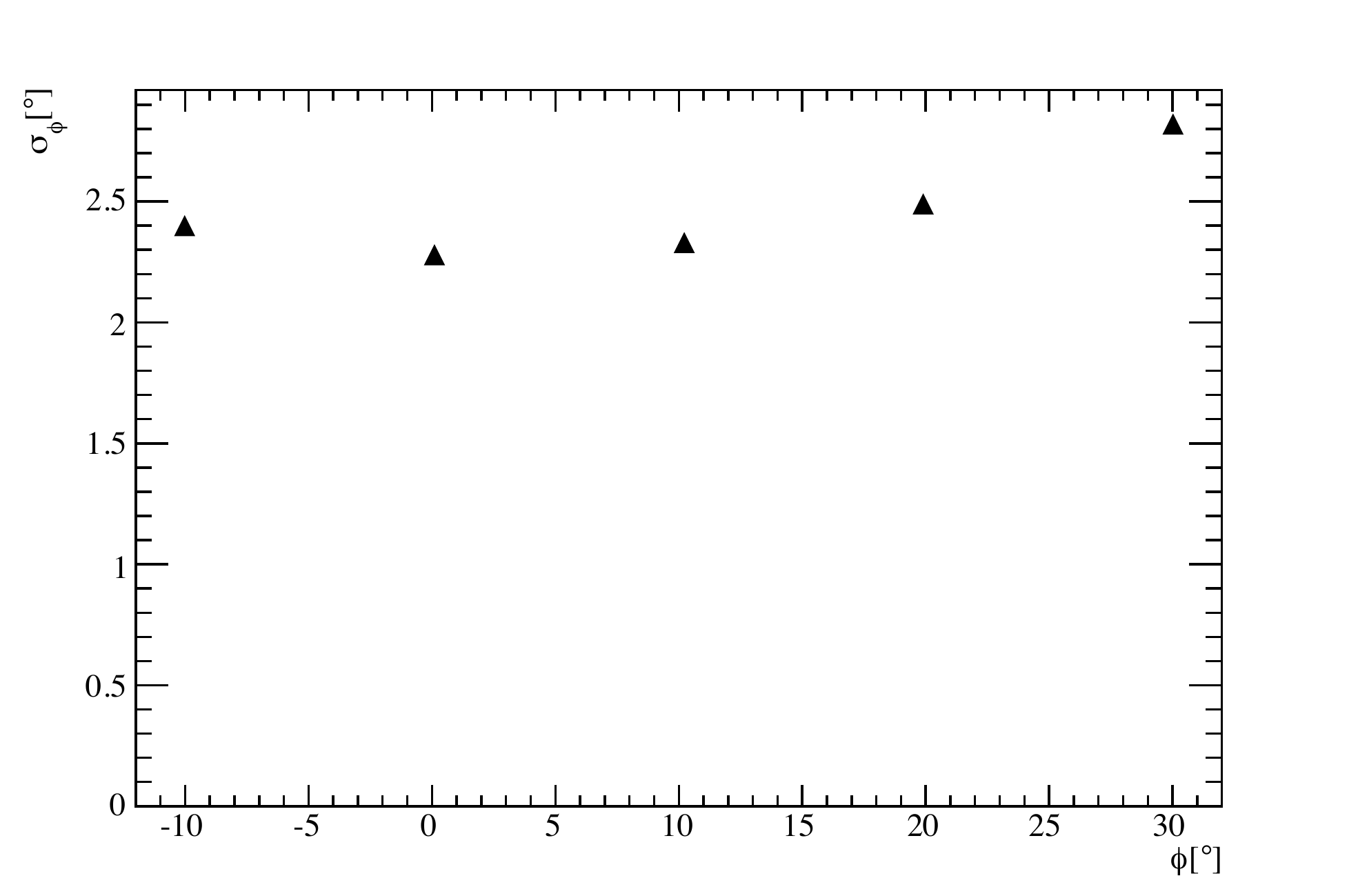}
}
  \caption{The uncertainty in the measured phase $\phi = \phi_{MIX} - 2\beta_c$ as a function of 
    the value of $\phi$ for $D^0\to \pi^+\pi^-$ decays at \lhcb with 5\invfb of data,
    assuming $\omega = \overline{\omega} = 0.06$.}
  \label{fig:lhcbprecision}
\end{center}
\end{figure}

Based on the result in Ref.~\cite{LHCb} we estimate that \lhcb will collect 
$7.8\e{6}$ $D^*$ tagged $\Dz\to \pi^+\pi^-$ decays in 5\invfb of data, based on an initial
37\invpb of data.  Based on the data shown in the reference, we estimate a purity of $\simeq 90\%$ and
$\omega \simeq 6\%$.  From these values, we obtain the sensitivities for $\arg(\lamf)\equiv\phi$ as a 
function of the phase shown in Fig.~\ref{fig:lhcbprecision} assuming this yield and mis-tag 
probability.

\subsection{Summary of sensitivity estimates}

Measurements of $\arg(\lamf)$ in all scenarios will require good knowledge of
the $\Dz\Dzb$ mixing parameters.  These should be available from SFF's running at \FourS and
from \lhcb.  \superb, for example, expects~\cite{O'Leary:2010af}
to measure these with precisions of a few times $10^{-4}$ for \xd and \yd, 
$\sim 1.5^{\circ}$ for $\phi_{MIX}$ and a few \% in $|q/p|$.  More information comes from 
the $D^0\to K^+K^-$ sample, and it is likely that \lhcb will further improve these parameters.  Hence, in a fit combining these measurements, it will be possible to separate out contributions from the mixing and weak phase 
in $D^0\to \pi^+\pi^-$ decays.  More accurately, the difference 
between phases of $\lamf$ measured in $D^0\to K^+K^-$ and $D^0\to \pi^+\pi^-$
decays is $\phi_{CP} =  -2\beta_{c,eff}$.  
If loop contributions can be well measured and both long-distance and weak 
exchange contributions are negligible, then this constraint can be translated into 
a measurement of $\beta_c$.  

In order to relate the measurement of the weak phase, $\beta_{c,eff}$ of \lamf to $\beta_c$, one needs
to measure a set of Isospin related $\D \to hh$ decays.  This is something that will require the \epem
environment, as it will not be possible for \lhcb to reconstruct $\Dz\to \piz\piz$, or $\D^0\to \rho^+\rho^-$. 
Both $\D^+\to \pi^+\piz$ and $\D^+\to \rho^+\rho^0$ would also be challenging measurements for \lhcb.  Nonetheless a 
search for \CP violation in $\D^0\to \pi^+\pi^-$ and $\D^0\to \rho^0\rho^0$ decays and a measurement of $\phi_{MIX}$ at \lhcb would be of considerable interest.

The corresponding sensitivity estimates for arg$(\lamf)$ in the 
different scenarios considered are summarized in Table~\ref{tbl:numericalanalysis}.  It should be recalled, however, that effects from time resolution, or from time-dependencies in efficiency or mis-tag rates are neglected here.  
We estimate that it should be possible to measure $\phi_{CP}$ to 
$\sim 2.6^\circ$ using this approach.  Assuming that penguin contributions can be measured precisely,
then the error on $\beta_c$ that could be obtained by \superb would be $\sim 1.3^\circ$.  \lhcb will
require input from \superb on the decay modes with neutral particles in the final state in order 
to translate a measurement of $\beta_{c,eff}$ to one on $\beta_c$.  
Further work is required to understand how penguins and other suppressed amplitudes 
affect the translation of $\beta_{c,eff}$ to $\beta_c$,
however it is clear that there will be a significant contribution from penguins given the
size of the $\Dz\to \piz\piz$ branching fraction.
It is worth noting that \babar and \belle could be able to make a measurement 
of $\beta_{c,eff}$ with a precision of $\sim 25^\circ$, using the nominal values
of $x$ and $y$ measured for charm mixing available today. We have highlighted several 
decays that could be used to measure this angle, including  $D\to \pi\pi$, $\rho\pi$, $\rho\rho$, $a_1\pi$, 
$K^0f^0$, and $K^0a^0$.  Ultimately it will be important to measure $\beta_{c,eff}$ in each of these
modes in order to cross-check the consistency of all of the measurements, constrain NP, and bound 
possible corrections to the CKM mechanism.

\begin{table}[!ht]
\caption{Summary of expected uncertainties from 500\invfb of data at charm threshold, 75\invab of 
data at the \FourS, and 5\invfb of data from \lhcb for $\Dz\to K^+ K^-$ and $\Dz\to\pi^+\pi^-$ decays.
The column marked SL corresponds to semi-leptonic tagged events, and 
the column SL+K corresponds to semi-leptonic and kaon tagged events at charm threshold.}\label{tbl:numericalanalysis}
\begin{center}
\begin{tabular}{l|ccc|c}
                          &  \multicolumn{3}{c|}{\superb} & \lhcb \\
Parameter                 & SL & $\,\,$SL + K $\,\,$ & \FourS &  \\ \hline \hline
$\phi(\pi\pi) = \arg(\lambda_{\pi\pi})$      & $8.0^\circ$           & $3.4^\circ$ & $2.2^\circ$ & $2.3^\circ$ \\
$\phi(KK) = \arg(\lambda_{KK})$      & $4.8^\circ$           & $2.1^\circ$ & $1.3^\circ$ & $1.4^\circ$ \\
$\phi_{CP}=\phi_{KK}-\phi_{\pi\pi}$ & $9.4^\circ$ & $3.9^\circ$ & $2.6^\circ$ & $2.7^\circ$ \\
$\beta_{c,eff}$           & $4.7^\circ$           & $2.0^\circ$ & $1.3^\circ$ & $1.4^\circ$ \\ \hline
\end{tabular}
\end{center}
\end{table}

Here we have concentrated on the determination of the phase $\arg(\lamf)$, and one should not neglect
the fact that we are also able to constrain $|\lamf|$ using these same measurements.  An observation of
$|\lamf| \neq 1$ in data would constitute the measurement of direct \CP violation in a given decay 
channel.  We estimate that it should be possible to measure $|\lamf|$ with a statistical uncertainty 
of $1-4\%$ at the future experiments discussed above, though we note that this is limited by any 
uncertainty in $\Delta\omega$.  This is smallest in \superb running at charm threshold, but is 
statistically limited less in other scenarios.

If one compares the relative power of data from charm threshold with that from the \FourS, it
is clear from Table~\ref{tbl:numericalanalysis} that 75 (50)\invab of data at the \FourS is equivalent
to approximately 1.2 (0.8)\invab at charm threshold.  It is interesting to note
that \superb proponents expect to accumulate 500\invfb of data at charm threshold in only
three months, whereas 75\invab would require five years of running at nominal luminosity.
The time-scale involved for the \belletwo experimental run at the \FourS is similar to the \superb
one.

\subsection{Constraint on the $cu$ triangle}
\label{sec:numerical:triangle}

It is possible to constrain the apex of the $cu$ triangle in Figure~\ref{fig:triangles} by constraining
two internal angles, or by measuring the sides.  If one considers the representation where the 
baseline $\vus^*\vcs$ is normalized to unity, then the angles at vertices corresponding to the 
coordinates $(0,0)$ and $(1,0)$ are $\beta_c$ and $\alpha_c$, respectively.  The constraint on the 
apex of the $cu$ triangle can be obtained using the CKM prediction of $\gamma_c = (68.4 \pm 0.1)^\circ$ (from the $B_d$ triangle), 
and any future measurement of $\beta_c$.  The $\gamma_c$ constraint is essentially a straight line in 
the complex plane containing the $cu$ triangle. As is the case with the $B_d$ triangle, there are multiple solutions for the apex
of the triangle. Even a rudimentary constraint on $\beta_c$, made by establishing that
$\beta_{c, eff}$ is compatible with zero, would constitute a test of the SM.
A precision measurement of $\beta_{c, eff}$ would require a detailed treatment of theoretical 
uncertainties to determine if any small deviation from the expected value of 
$\beta_c$ was due to new physics, or compatible with the SM.
This is an area that will require work in the future.  We are currently working on
determining the effect of penguin pollution in $D\to hh$ decays.  In addition to
this effect, other potential sources of theoretical uncertainty that may be relevant
include isospin breaking effects, long distance topologies or failure of the factorization hypothesis.
The coordinates of the apex of the triangle are given by
\begin{eqnarray}
X + iY&=& 1 + \frac{A^2 \lambda^5 (\overline{\rho} + i \overline{\eta})}{\lambda - \lambda^3/2 - \lambda^5(1/8 + A^2/2)},
\end{eqnarray}
neglecting contributions from all higher orders in $\lambda$.  Given that the apex of the $bd$ triangle
is $\overline{\rho} + i \overline{\eta}$, one can over constrain the SM by testing 
the prediction of $X+iY$ from existing constraints on the apex of the $bd$ triangles.  We find that 
\begin{eqnarray}
 X &=& 1.00025, \\
 Y &=& 0.00062,
\end{eqnarray}
using the existing constraints on the Wolfenstein parameters.  


In order to measure $\beta_{c,eff}$ one needs to precisely constrain $\phi_{MIX}$.
The current method to measure the mixing phase is via a time-dependent 
Dalitz Plot analysis of $D$ decays to self conjugate final states.  Here we
propose to use a time-dependent analysis of decays such as $D\to K^+K^-$, which have an
overall phase dominated by the mixing phase in the SM assuming the CKM 
parameterization, and rate larger than the $\pi\pi$ channel.  Having 
determined $\phi_{MIX}$ one can then
decouple the mixing phase contribution in $D\to \pi^+\pi^-$ decays, 
and by performing an Isospin analysis one can translate a measurement
of \lamf into a constraint on $\beta_{c,eff}$.  Alternatively one can use a 
model independent measurement of the mixing phase, to decouple
$\phi_{MIX}$ and $\beta_{c,eff}$ from the measurement of \lamf.
One would have 
to control both theoretical and systematic uncertainties to below one per mille in order to be sure 
of measuring a non-zero value of $\beta_c$.  At this time it is unclear if this 
will be achievable, however the \superb experiment has the added advantage of
being able to study the time-dependence in two ways, and hence may be able to 
avoid limitations inherent to the $\D^*$ tagged analyses.

\section{$\B_{d}$ decays}
\label{sec:bdecays}

The effect of a non-zero $\Delta \Gamma$ on the time-dependent \CP asymmetry distribution
is an alteration of the phase of oscillation, and of the amplitude of the oscillation as a function
of $t$ or \deltat.  Until now all time-dependent \CP asymmetry measurements in $B_d$ decays 
have assumed that $\Delta \Gamma = 0$, which was a reasonable assumption based on theoretical
expectations.  However it should be noted that it is possible to bound the systematic 
uncertainty in the measurement of the unitarity triangle angles $\alpha$ and $\beta$ by making 
this assumption using the known experimental constraint on 
${\mathrm sign}(Re\lamf) \Delta \Gamma/\Gamma = 0.010\pm 0.037$~\cite{pdg}.
If one compares the asymmetry obtained assuming $\Delta \Gamma$ corresponding to
the experimental bound, then it is possible to 
estimate the bias and systematic uncertainty time-dependent asymmetry measurements made in 
\B decays arising from the assumption that $\Delta \Gamma = 0$.

\begin{figure}[!ht]
\begin{center}
  \resizebox{8cm}{!}{\includegraphics{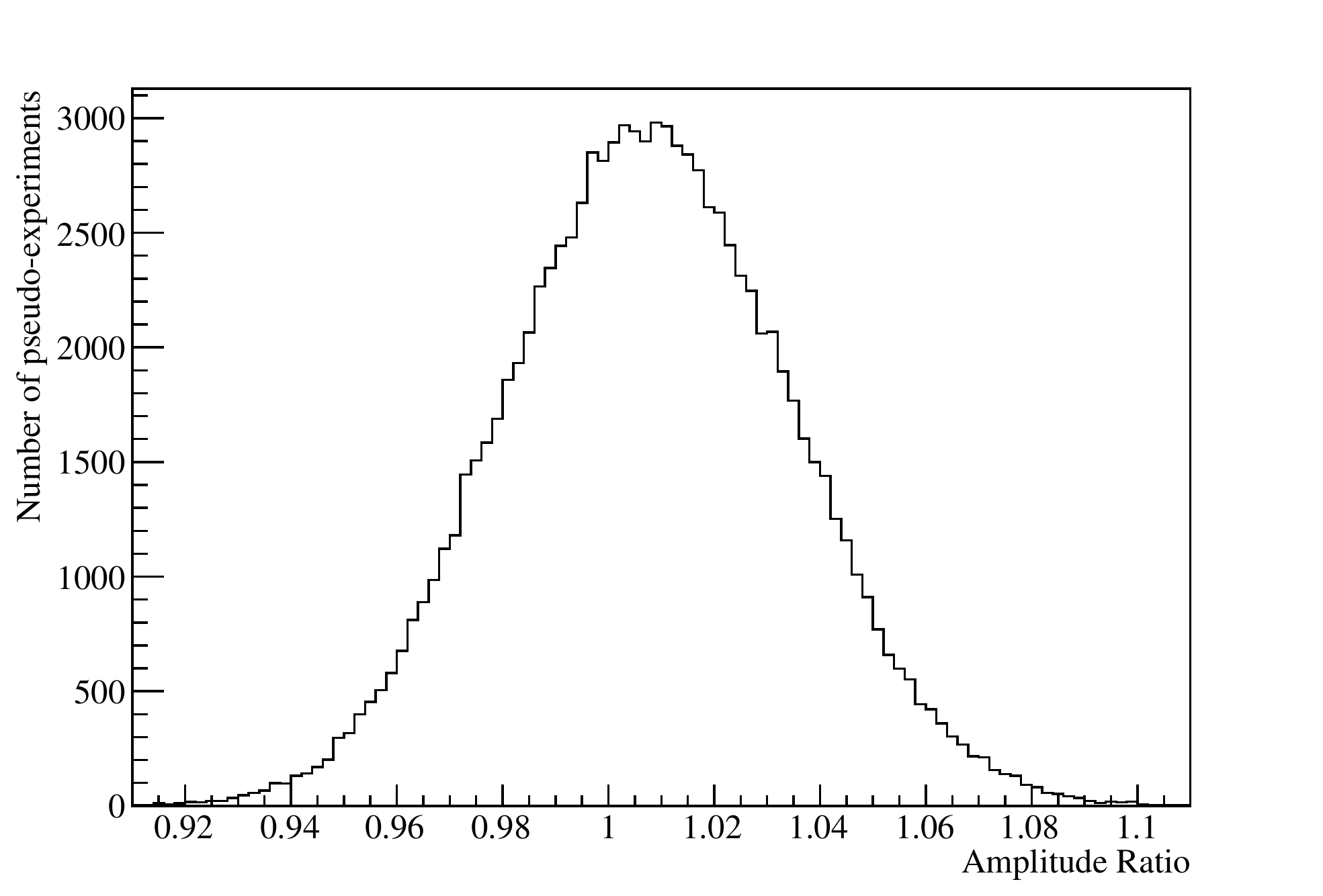}}
  \caption{The bias on $S=\sin2\beta$ obtained from a Monte Carlo based simulation 
  resulting from the assumption that $\Delta \Gamma = 0$.  The amplitude ratio plotted
  is that of the maximum time-dependent amplitude accounting for a non-zero $\Delta \Gamma$ 
  to that where $\Delta \Gamma=0$.}
  \label{fig:biasonsintwobeta}
\end{center}
\end{figure}

We have performed a Monte Carlo based simulation for the scenario of $S=0.7$ and $C=0.0$ 
taking the uncertainty in $\Delta \Gamma$ to be Gaussian.  The ratio of amplitudes for the 
first maximum/minimum obtained as an estimate of the systematic effect on $S=\sin2\beta$ is $0.007\pm 0.027$,
and the corresponding distribution is shown in Fig.~\ref{fig:biasonsintwobeta}.
This is comparable to the statistical uncertainty in $\sin 2\beta$ measurements~\cite{Aubert:2009yr,Chen:2006nk}.

Moving onto the measurements related to $\alpha$, if one considers $\Bz\to\pi^+\pi^-$ decays, 
where $S=-0.65\pm 0.07$ and $C=-0.38\pm 0.06$~\cite{Ishino:2006if,Aubert:2008sb}, then the systematic uncertainty in the 
measurement of $S$ and $C$ is $0.009\pm 0.032$.  In this case, the systematic effect resulting
from the assumption that $\Delta \Gamma=0$ is also non-trivial, but does not dominate the total
uncertainty.  The most important channel for the constraint on $\alpha$ is however 
$\Bz\to\rho^+\rho^-$ where $S=-0.05\pm 0.17$ and $C=-0.06\pm 0.13$~\cite{Aubert:rhorho,Somov:rhorho}.  The
corresponding systematic effect on $S$ and $C$ is $-0.008\pm 0.038$, which is currently small
compared to the experimental determination of those quantities.  Therefore, while the 
$\Delta \Gamma=0$ bound may impact upon the $\beta$ constraint imposed on the unitarity
triangle, it will have little effect on the measurement of $\alpha$.

Therefore current and future experiments aimed at performing a precision measurement of time-dependent 
\CP asymmetries should also strive to increase the precision of the bound on $\Delta \Gamma$ to ensure that 
this systematic effect does not dominate future measurements.

\section{$\B_{s}$ decays}
\label{sec:bsdecays}

Oscillations in $B_s$ decays are extremely fast relative to $B_d$ and $D$ mesons,
and so neither \superb or \belletwo are expected to be able to perform time-dependent
asymmetry measurements in $B_s$ decays.  It should however be noted that if these
experiments were to accumulate large samples of events at the \FiveS, then
the distribution of events as a function of $\deltat$ would contain information
on both the real and imaginary parts of $\lamf$.  Hence some information from CP
asymmetries related to the time-dependent measurements being done at hadron collider
experiments would be measurable in an \epem environment.  This was also discussed 
in~\cite{Baracchini:2007ei}
in the context of measurements of $B_s \to J/\psi \phi$ at \superb. This issue
is particularly relevant for final states including neutral particles such as 
$B_s \to \eta^\prime \phi$, the $B_s$ equivalent to the most precisely measured 
golden $b\to s$ penguin mode $\Bz \to \eta^\prime \Kz$.
It would be extremely challenging to study this mode a hadronic environment
and so the best way to study \CP violation in this mode would be using 
data collected at the \FiveS.

Other interesting decays to study are $B_s\to \rho\KS, D_s^\pm K^\mp$, and $B_s\to
D\phi$ as these 
measure $\gamma$~\cite{flesicher,aleksan,gronau}.  It would be interesting to compare the values
obtained from a $B_s$ decay with the result from the $B_d\to DK$ approach 
currently being used by experiments.  It should be noted that \lhcb should be
able to perform time-dependent measurements of these modes.  Finally, as noted
in Ref.~\cite{pdg}, the channel $B_s\to \piz\KS$ is equivalent to the 
channel $B_d\to \pi^+\pi^-$.  Therefore it would be interesting to attempt
to measure $\lamf$ for this decay.  Given the $\piz$ in the final state, 
and lack of information to constrain a primary vertex, this could be an excellent
candidate for \superb or \belletwo to study.

\section{Conclusions}
\label{sec:conclusions}

We have outlined the formalism required to 
experimentally measure time-dependent \CP asymmetries in charm decays
using correlated $\Dz\Dzb$ decays as well as $\Dz$ mesons tagged from $D^*$
decays, and discussed the benefits of studying a number of 
different \CP eigenstates.  The important points to note are that one can use
$K^+K^-$ decays to measure the mixing phase quite precisely and other decays can be used to constrain the angle $\beta_{c,eff}$ which is related to the $cu$ unitarity triangle.  
These observables are also sensitive to possible enhancements from new physics.
A data sample of 500\invfb collected at charm threshold would provide a 
sufficient test to constrain any potential large NP effects.
Similar measurements would also be possible using $D^*$ tagged decays at \superb, \belletwo and 
\lhcb.  From event yields currently available, we expect the statistical precision in the measured 
phase at \superb to be slightly better than results from a 5 \invfb \lhcb run. 
As the $cu$ and the $bd$ unitarity triangles are related, the measurements
proposed here provide a new set of consistency checks on the unitarity of the CKM matrix that can be performed using $D$ decays.  
Measurements of the sides of these triangles would
enable a further, indirect cross-check on the validity of this matrix.  Only the \superb experiment will be able to make a complete set of the measurements
required to perform direct and indirect constraints of both triangles.  As $\beta_c$ is an extremely small angle, its determination will 
be limited by theoretical and systematic uncertainties.  \superb has a potential advantage over other experiments as it will be able to collect data at charm threshold with a boosted center of mass, as well as being able to explore effects using neutral mesons from $D^*$ tagged events.  Data from charm threshold will be almost pure, with a mis-tag probability of $\sim 0$ for semi-leptonic tagged events, which could be advantageous if systematic uncertainties dominate measurements from \FourS data and from \lhcb.  The ultimate theoretical uncertainty in relating $\beta_{c,eff}$ to $\beta_c$ needs to be evaluated.  
A measurement of $| \lamf | \neq 1$ could also signify direct \CPV.

We also point out that precision measurements of time-dependent asymmetries
in $B_d$ decays require improvements in our knowledge of $\Delta\Gamma_{\B_d}$.
The current experimental constraint on this observable translates into a
systematic effect of the order of $0.007\pm 0.027$, which is comparable with 
the current experimental sensitivity on $\sin 2\beta$ from \babar and \belle.
We have also computed the systematic effect of assuming $\Delta\Gamma_{\B_d}=0$ 
for measurements of $\alpha$ from $\Bz\to\pi^+\pi^-$ and $\Bz\to\rho^+\rho^-$ decays,
which is negligible for existing measurements.

It may be possible to measure the real and imaginary parts of \lamf
from a simplified time-dependent analysis of $B_s$ decays at \superb and 
\belletwo without the need to observe oscillations. While the approach outlined 
would not be competitive with modes that could be measured in a hadronic 
environment, it would provide unique access to observable channels that would 
be inaccessible to the Tevatron and \lhcb.  The prime example is that 
of $B_s \to \eta^\prime \phi$, which is the direct analog of the most precisely
measured $B_d^0\to s$ penguin mode $B_d^0 \to \eta^\prime \Kz$ from the \B factories.

\section{Acknowledgments}

This work has been supported by the US National Science Foundation,
under grant number PHY-0757876, and G. Inguglia received financial
support from Queen Mary, University of London during the preparation of this paper.
The authors would like to thank Marco Ciuchini and Matteo Rama for useful 
comments on this paper.

\bibliography{note}



\end{document}